\documentclass[aps,preprint,nofootinbib]{revtex4}
\usepackage{amsmath}
\usepackage{amssymb}
\usepackage{verbatim}
\begin{document}
\author{R. N. Ghalati}
\email{rnowbakh@uwo.ca}
\affiliation{Department of Applied Mathematics,
University of Western Ontario, London, N6A~5B7 Canada}
\title{
A Novel Hamiltonian Formulation of First Order Einstein-Hilbert
Action: Connection with ADM, Diffeomorphism Invariance and Linearized Theory}
\date{\today}
\preprint{{\footnotesize UWO\,-TH-\,09/1}}
\begin{abstract}
A novel Dirac Hamiltonian formulation of the first order Einstein-Hilbert (EH)
action, in which ``algebraic'' constraints are not solved to eliminate fields from the
action at the Lagrangian level, has been
shown to lead to an action and a constraint structure apparently distinct from the
ADM action and the ADM constraint structure in
that secondary first class constraints $\chi$ and $\chi_i$ as well as
tertiary first class constraints $\tau$ and $\tau_i$ arise with an
unusual Poisson Bracket (PB) algebra \cite{5Ghalati2007-2}. By canonical transformations of the
fundamental fields we show how from the tertiary constraints $\tau$
and $\tau_i$ one may derive the Hamiltonian and momentum constraints. Special
attention is paid to the Hamiltonian formulation of the first order EH
action in terms of the
variables $h=\sqrt{-g}g^{00}$, $h^i=\sqrt{-g}g^{0i}$ and
$q^{ij}=-g\,(g^{0i}g^{0j}-g^{00}g^{ij})$ and their conjugate momenta
employed in \cite{5Faddeev1975,5Faddeev1982}.  
It is shown that the variables $h$ and $h^i$ are left undetermined in
the formalism. This fact is used for a proper gauge fixation of the
secondary constraints $\chi$ and $\chi_i$ and reduction to the Faddeev
action \cite{5Faddeev1975,5Faddeev1982}.  Considering invariances
of the total action, the generator of the gauge
transformations of the EH Lagrangian action is derived. Using this generator,
the explicit form of the gauge invariance of the field $h$ is
obtained, by which the relation between the gauge functions and the
descriptors of the diffeomorphism invariance is determined in order
for the gauge transformations to correspond to diffeomorphism invariance. By linearizing
the novel Hamiltonian formulation of \cite{5Ghalati2007-2}, the
Hamiltonian formulation of the first order action for the free spin
two field \cite{5Arnowitt1959-1,5Ghalati2007-2} is derived. 
\end{abstract}
\maketitle
\section{Introduction}
After the discovery of the Dirac constraint formalism
\cite{5Anderson,5Dirac1950}, the Hamiltonian formulation of the EH
action in second order form using the metric $g_{\mu\nu}$ as
the configuration space fields was first attempted by 
Pirani and Schild \cite{5Pirani1950,5Pirani1952} and independently by
Bergmann, Penfield, Schiller and Zatzkis \cite{5Bergmann1950}, and was
later formulated in a more convenient way by Dirac
\cite{5Dirac1958,5Dirac1959}. Soon after wards, the canonical formulation of the 
first order Einstein-Palatini action was considered by
Arnowitt, Deser and Misner (ADM)
\cite{5Arnowitt1959-1,5Arnowitt1959-2,5Arnowitt1960,5Arnowitt1960-2,5Arnowitt}
starting from a geometrical, rather than an algebraic perspective. In
achieving their formulation, ADM followed a procedure other than the  
Dirac constraint formalism, in which constraint equations are solved
irrespective of their being first or second class
\cite{5Faddeev1988}. The result is derived using a set of variables 
possessing clear geometrical interpretation. It turns out that in both the Dirac and
ADM formulations, the metric of the three-space and their
conjugate momenta which are related to the extrinsic curvature of the
three-space (subject to the Hamiltonian and momentum
constraints), are sufficient for the description of the 
dynamics of general relativity, which is considered as the time
evolution of spacelike surfaces. A characteristic of both
formulations is that the manifest four dimensional general covariance is
broken, which is to be expected by the choice of a
particular time coordinate necessary for the Hamiltonian
formulation. In early attempts, however, care was taken for a
canonical formulation in terms of invariants \cite{5Pirani1950,5Bergmann1950}, but this was soon
overshadowed by abandoning such assumptions \cite{5Pirani1952}, and especially
after Dirac's triumphant results \cite{5Dirac1958,5Dirac1959}.

A key element in the ADM Hamiltonian formulation of the EH action
in first order form is the ``reduction'' of the EH action by solving a
combination of equations of motion which are independent of time derivatives (the algebraic
constraints), thus eliminating a number of dynamical variables from
the Lagrangian action. This algebraic manipulation, which brings the EH action
in a Hamiltonian form \cite{5Arnowitt1960-2,5Faddeev1982}, is done 
irrespective of whether the equations of motion which are solved are
first or second class in the sense of the Dirac constraint formalism
\cite{5Faddeev1988}. Such a formalism has thus left untouched 
the question of what kind of a Hamiltonian formulation, with what
characteristics, and potentially what differences, one would have obtained if one
had used the Dirac constraint formalism in when
casting the EH action in the Hamiltonian form. This task was recently
undertaken in \cite{5Ghalati2007-2}. 

Here is a brief sketch of this paper. A summary of the ADM approach,
in its original formulation
\cite{5Arnowitt1959-1,5Arnowitt1959-2,5Arnowitt1960,5Arnowitt1960-2,5Arnowitt}
and the formulation of Faddeev \cite{5Faddeev1975,5Faddeev1982}, as well as an
overview of the results of the novel Hamiltonian formulation of  
\cite{5Ghalati2007-2} are discussed in Section (\ref{I}). In
sections (\ref{II}) and (\ref{III}) we explain how one may simplify the
form of the constraints and the constraint algebra appearing in
\cite{5Ghalati2007-2} by transforming the coordinates employed in
\cite{5Ghalati2007-2} into the variables used
by Faddeev \cite{5Faddeev1975,5Faddeev1982}, ADM
\cite{5Arnowitt1959-2,5Arnowitt1960,5Arnowitt1960-2,5Arnowitt1960} and Teitelboim
\cite{5Teitelboim1982,5Teitelboim1983}.
It is then shown how one may reduce these actions into the actions
derived by Faddeev \cite{5Faddeev1975,5Faddeev1982} and ADM
\cite{5Arnowitt1959-2,5Arnowitt1960,5Arnowitt1960-2,5Arnowitt} using the
method of Faddeev and Jackiw \cite{5Faddeev1988}. Based on the
equations of motion for $h$ and $h^i$, when $h$, $h^i$ and $q^{ij}$ are
used as coordinates, tentative gauge constraints are suggested for
reduction of the extended action into the Faddeev action
\cite{5Faddeev1975,5Faddeev1982} in Section (\ref{iii}).  Gauge
invariance of this action is considered in Section (\ref{ii}), where the
generator f the gauge transformations of the total action is
derived. Using this generator, the explicit form of the gauge
transformation of the field $h$ is obtained, and the relation between
the gauge functions and the descriptors of the diffeomorphism
invariance is determined for the gauge transformation to be a diffeomorphism.  
In Section (\ref{XV}), the
linearized form of the Hamiltonian formulation of the EH action of ref.
\cite{5Ghalati2007-2}, which is the Hamiltonian action corresponding to the first
order spin two field Lagrangian action proposed in
\cite{5Arnowitt1959-1}, is obtained. Concluding 
remarks are left to Section (\ref{IV}).

\section{Summary of ADM approach and previous results}\label{I}
ADM achieved their Hamiltonian formulation of the EH action by
casting it in the form \cite{5Arnowitt1959-2,5Arnowitt1960,5Arnowitt1960-2}
\begin{eqnarray}\label{ghd4}
\sqrt{-\mathfrak{g}}\,R&\simeq&\mathcal{L}\left(N,N_i,\gamma_{ij}\right)\\
\nonumber &=& -\,\gamma_{ij}\,  
\dot{\pi}^{ij}+\frac{N}{\sqrt{\gamma}}\left(\gamma
\,\mathcal{R}+\frac{1}{2}\left(\pi^l_{\,l}\right)^2-\pi^{ij}\pi_{ij}\right)+N_i\left(2\,
\pi^{ij}_{\,\,\,\,|j}\right)\\
\nonumber  
&-&2\left[\sqrt{\gamma}\,N_{,\,i}+\left(\pi_{ij}-\frac{1}{2}\,\gamma_{ij}\,\pi^l_{\,l}\right)N^j\right]^{|i}\,,
\end{eqnarray} 
where $\gamma=\det(\gamma_{ij})$,
\begin{equation}\label{ghd5}
\pi^{ij}=\sqrt{-\mathfrak{g}}\left[\Gamma^0_{mn}-\gamma_{mn}\Gamma^0_{pq}\gamma^{pq}\right]\gamma^{mi}\gamma^{nj},  
\end{equation}
and
\begin{equation}\label{ghd6}
N=\left(-g^{00}\right)^{-1/2}\,\,\,\,\,\,\,\,\,\,\,\,\,\,N_i=g_{0i}.
\end{equation}
$N$ and $N_i$ are treated as configuration fields instead of $g_{00}$ and
$g_{0i}$, spanning the configuration space together with the metric of
the 3-space $\gamma_{ij}$. In eq. (\ref{ghd4}), $\mathcal{R}$ is the
curvature scalar of the $3$-space $\gamma_{ij}$, and the vertical dash $|$ denotes covariant
derivative with respect to the 3-space $\gamma_{ij}$ defined
as usual. In particular, if $\mathcal{S}$, $\mathcal{T}^{ij}$ and
$\mathcal{T}^i_{\,\,j}$ are tensor densities of rank $W$ we have \cite{5Carmeli,5Weinberg}
\begin{eqnarray}\label{covariantderiv0}
\mathcal{S}_{|k}&=&\mathcal{S}_{,k}-W\,\Gamma^l_{lk}\,\mathcal{S}\,,\\ 
\label{covariantderiv1} 
\mathcal{T}^{ij}_{\,\,\,\,\,|k}&=&\mathcal{T}^{ij}_{\,\,\,\,,k}+\Gamma^i_{kl}\,\mathcal{T}^{lj}+\Gamma^j_{kl}\,\mathcal{T}^{il}-W\,\Gamma^l_{lk}\,\mathcal{T}^{ij}\,,\\
\label{covariantderiv2}  
\mathcal{T}^i_{\,\,\,j\,|k}&=&\mathcal{T}^i_{\,\,\,j,k}+\Gamma^i_{kl}\,\mathcal{T}^l_{\,\,\,j}-\Gamma^l_{jk}\,\mathcal{T}^i_{\,\,\,l}-W\,\Gamma^l_{lk}\,\mathcal{T}^i_{\,\,\,j}\,.
\end{eqnarray}
  
The action of
eq. (\ref{ghd4}) is obtained from the first order EH action by solving
linear combinations of the equations of motion derived from this action, which are
independent of the time derivative of fields (the constraint
equations), for the components $\Gamma^i_{jk}$, $\Gamma^i_{0k}$ and $\Gamma^0_{0k}$ of
the affine connections in terms of the lapse and shift functions $N$
and $N_i$, the metric fields of the 3-space $\gamma_{ij}$ and the
components $\Gamma^0_{ij}$ of the affine 
connections, and by dropping some surface integrals. This is done without classifying the
constraints. The notation $\simeq$ rather than $=$ is used in
eq. (\ref{ghd4}) since the equality holds only if these solutions to
the equations of motion are substituted into $R$. (The components
$\Gamma^\mu_{00}$ of the affine connections disappear from the action
when the solutions for the constraint equations are 
inserted and are not considered by ADM.)
 
The term appearing in the total divergence in eq. (\ref{ghd4}) is a covariant vector density
of weight $W=1$. The total divergence may be dropped (as it is done
below) if compact spaces are under consideration. From
eq. (\ref{ghd4}), we see that $\pi^{ij}$, which is a contravariant tensor density of
weight $W=1$, is the momenta conjugate to
$\gamma_{ij}$. Therefore, the canonical Hamiltonian corresponding to
the action of eq. (\ref{ghd4}) is given by
\begin{equation}\label{ghd7}
H_{ADM}=\int dx \left(N\mathcal{H}+N_i\mathcal{H}^i\right)\,,
\end{equation}
where
\begin{eqnarray}\label{ghd8}
\mathcal{H}&=&\gamma^{-1/2}\left(\pi^{ij}\pi_{ij}-\frac{1}{2}\left(\pi^l_{\,l}\right)^2\right)-\gamma^{1/2}\,\mathcal{R}\,,  
\\ \label{ghd8-2}   
\mathcal{H}^i&=&-2\,\pi^{ij}_{\,\,\,\,|j}=-2\left(\pi^{ij}_{\,\,,j}+\Gamma^i_{jk}\,\pi^{jk}\right) \,,
\end{eqnarray}
are called the ``Hamiltonian'' and ``momentum'' constraints
respectively. The nomenclature becomes clear in the following
way. Variation of the action with respect to the fields $N$ and $N^i$,
which act as Lagrange multiplier fields, gives rise to the constraints
\begin{equation}\label{3secon}
\mathcal{H}\approx0\,\,\,\,\,\,\,\,\,\,\,\,\,\,\,\,\,\,\,\,\,\,\,\,\,\,\,\,\,\,\,\,\,\,\,\,\,\,\,\mathcal{H}_i\approx0\,.      
\end{equation}  
These constraints satisfy the algebra \cite{5Sundermeyer}
\begin{eqnarray}\label{3admalgebra}
\left\{\mathcal{H}(x),\mathcal{H}(y)\right\}&=&\left[\gamma^{ij}\mathcal{H}_j(x)+\gamma^{ij}\mathcal{H}_j(y)\right]\partial_i\delta(x-y)\,,\\
\nonumber
\left\{\mathcal{H}_i(x),\mathcal{H}(y)\right\}&=&\mathcal{H}(x)\,\partial_i\delta(x-y)\,\\
\nonumber 
\left\{\mathcal{H}_i(x),\mathcal{H}_j(y)\right\}&=&\left[\mathcal{H}_j(x)\,\partial_i+\mathcal{H}_i(y)\,\partial_j\right]\delta(x-y)\,,
\end{eqnarray}
which implies that the time change of the constraints
$\mathcal{H}$ and $\mathcal{H}_i$ is ensured to weakly vanish when
computed using the ADM Hamiltonian of eq. (\ref{ghd7}). The PBs
of the ADM canonical coordinates $\gamma_{ij}$ with the Hamiltonian
and momentum constraints $\mathcal{H}$ and $\mathcal{H}_i$ have neat
interpretations \cite{5Misner,5Sundermeyer}; namely, 
\begin{eqnarray}\label{3cov}
\left\{\gamma_{ij}\,, \int dx\,
N^k\mathcal{H}_k\right\}&=&\gamma_{ij,k}\,N^k+\gamma_{ik}\,N^k_{,j}+\gamma_{jk}\,N^k_{,i}\\
\nonumber
&=&N_{i|j}+N_{j|i}
\end{eqnarray} 
is nothing but the diffeomorphism invariance of the metric components $\gamma_{ij}$
of the spacelike surfaces, and
\begin{eqnarray}\label{3cov2}
\left\{\gamma_{ij}\,, \int dx\,
N\mathcal{H}\right\}&=&2\,N\,\gamma^{-1/2}\left(\,\pi_{ij}-\frac{1}{2}\,\pi^l_{\,l}\,\gamma_{ij}\right)\,,
\end{eqnarray}  
when set to zero, is the dynamical equation for the metric components
$\gamma_{ij}$ \cite{5Arnowitt}; thus the 
nomenclature for the ``Hamiltonian'' and ``momentum'' constraints
$\mathcal{H}$ and $\mathcal{H}_i$.   

Having reviewed the original Hamiltonian formulation of ADM
\cite{5Arnowitt1959-1,5Arnowitt1959-2,5Arnowitt1960,5Arnowitt1960-2,5Arnowitt},
we note that for the Hamiltonian formulation one may choose to start
with the EH action written in terms of the metric density
$h^{\mu\nu}=\sqrt{-\mathfrak{g}}\,g^{\mu\nu}$ and the affine connection 
$\Gamma^\lambda_{\mu\nu}$ as independent fields,
\begin{equation}\label{3ax3}
S=S\left(h^{\mu\nu},\Gamma^\lambda_{\mu\nu}\right)=\int dx\,
  h^{\mu\nu} \left(\Gamma^\lambda_{\mu \nu ,
  \lambda}-\Gamma^\lambda_{\lambda \mu , \nu} + \Gamma^\lambda_{\mu
  \nu} 
\Gamma^\sigma_{\sigma \lambda}-\Gamma^\lambda_{\sigma \mu} \Gamma^\sigma_{\lambda \nu}\right)\,.
\end{equation}
This choice of variables is made in
\cite{5Faddeev1975,5Faddeev1982}. An advantage of such a choice is
that it eliminates the square root of the determinant of the metric of the
3-space in the final Hamiltonian 
formulation; terms including this factor appear in the Hamiltonian 
constraint of eq. (\ref{ghd8}). A similar set of variables have been employed in 
the novel Dirac Hamiltonian formulation of the first order EH action
presented in the following subsection. A total divergence
appears in equations below which corresponds to the fact that in
\cite{5Faddeev1975,5Faddeev1982} asymptotically flat spacetimes,
rather than closed spacetimes, have been considered \cite{5Teitelboim1973,5Wald}.

The EH action of eq. (\ref{3ax3}), after addition of a surface term, becomes
\begin{equation}\label{3ax2}
S\left(h^{\mu\nu},\Gamma^\lambda_{\mu\nu}\right)=\int dx\,
  \left[\Gamma^\sigma_{\nu\sigma}\,h^{\mu\nu}_{\,\,,\mu}-\Gamma^\sigma_{\mu\nu}\,h^{\mu\nu}_{\,\,,\sigma} 
  +h^{\mu\nu}\left( \Gamma^\lambda_{\mu 
  \nu} 
\Gamma^\sigma_{\sigma \lambda}-\Gamma^\lambda_{\sigma \mu}
  \Gamma^\sigma_{\lambda \nu}\right)\right]\,. 
\end{equation}
Thirty of the equations of motion that arise from the action of
eq. (\ref{3ax2}) are independent of time derivatives and can be written as
\begin{eqnarray}\label{msm1}
h^{ik}_{\,\,,\,l}+\Gamma^i_{l\sigma}\,h^{\sigma
  k}+\Gamma^k_{l\sigma}\,h^{\sigma
  i}-h^{ik}\Gamma^\sigma_{l\sigma}&=&0\,,\\ \label{msm2} 
h^{0k}_{\,\,,\,l}+\Gamma^0_{lm}h^{mk}+\Gamma^k_{lm}h^{0m}+\Gamma^k_{l0}h^{00}-h^{0k}\Gamma^m_{lm}&=&0\,,\\\label{msm3}
h^{00}_{\,\,,\,l}+2\Gamma^0_{lm}h^{m0}+\Gamma^0_{l0}h^{00}-h^{00}\Gamma^m_{lm}&=&0\,.
\end{eqnarray}
These equations are used by Faddeev \cite {5Faddeev1975,5Faddeev1982}
to eliminate the variables $\Gamma^l_{ik}$, 
$\Gamma^k_{i0}$ and $\Gamma^0_{i0}$ from the action of
eq. (\ref{3ax2}). The reduced action is
\begin{eqnarray}\label{5Faddeev}
S_R=\int dx\, \left[\,\Pi_{ik}\,\dot q^{ik}-\lambda^0C_0-\lambda^kC_k-\mathcal{H}\right]\,, 
\end{eqnarray} 
where
\begin{eqnarray}\label{mnm1}
C_k&=&2\bigtriangledown_k(q^{il}\Pi_{il})-2\bigtriangledown_l(q^{il}\Pi_{ik})\,,\\\label{mnm2}
C_0&=&q^{ik}q^{mn}(\Pi_{ik}\Pi_{mn}-\Pi_{im}\Pi_{kn})+\gamma
\,\mathcal{R}\,\\ \label{mnm3}
\mathcal{H}&=&-C_0-q^{ik}_{\,\,,ik}\,. 
\end{eqnarray}
In the above equations $q^{ik}=h^{0i}h^{0k}-h^{00}h^{ik}$ is a
contravariant metric density of weight $W=2$,
$\Pi_{ik}=\Gamma^0_{ik}/h^{00}$ is a covariant tensor density of weight $W=-1$,
the fields $\lambda^0=1+1/h^{00}$ and 
$\lambda^k=h^{0k}/h^{00}$ are Lagrange multiplier fields, $\bigtriangledown$ is the covariant
derivative with respect to the metric $\gamma_{ik}$ of the three
dimensional space as defined in
eqs. (\ref{covariantderiv0}-\ref{covariantderiv2}), $\mathcal{R}$ is its
scalar curvature and $\gamma=\det{(\gamma_{jk})}$. (Note that the
quantities $q^{il}\Pi_{il}$ and $q^{il}\Pi_{ik}$ appearing in
eq. (\ref{mnm1}) are scalar and mixed
second rank tensor densities of weight $W=1$ respectively.) The fields $\Gamma^\mu_{00}$ enter
linearly in eq. (\ref{3ax2}) and disappear in the reduced action; they
are no longer considered when counting degrees of freedom. At this
stage the only dynamical fields are $q^{ik}$ and their
conjugate momenta $\Pi_{ik}$. The fields $h^{00}$ and $h^{0k}$ are taken to be
non-dynamical in $S_R$ on the account of their appearing as Lagrange multiplier fields
through $\lambda^0$ and $\lambda^l$. Variation of the action with
respect to these Lagrange multipliers in turn results in the
constraints $C_0\approx0$ and $C_k\approx0$. The PBs of these
constraints are convenient to express in terms of the 
functionals \cite{5Faddeev1982}
\begin{eqnarray}\label{3funct-1}
C({\bf{X}})&=& \int C_k(x)X^k(x)\,dx\,,\\ \label{3funct-2}
C_0(f)&=&\int C_{0}(x)f(x)\,dx\,,
\end{eqnarray}
and are
\begin{eqnarray}\label{3com-1}
\big\{C({\bf{X}}_1)\,,C({\bf{X}}_2)\big\}&=&C\left(\left[{\bf{X}}_1,{\bf{X}}_2\right]\right)\,,\\
\label{3com-2}
\big\{C({\bf{X}})\,,C_0(f)\big\}&=&C_0\left({\bf{X}}f\right)\,,\\ \label{3com-3}
\big\{C_0(f_1)\,,C_0(f_2)\big\}&=&C\left(\left[f_1,f_2\right]\right)\,,
\end{eqnarray}
where 
\begin{eqnarray}\label{3def-1}
\left[{\bf{X}}_1\,,{\bf{X}}_2\right]^k&=&X^l_1\,\partial_lX^k_2-X^l_2\,\partial_lX^k_1\,,\\
\label{3def-2}
{\bf{X}}f&=&X^l\partial_lf-f\,\partial_lX^l\,,\\ \label{3def-3}
\left[f_1\,,f_2\right]^k&=&q^{ik}\left(f_1\partial_if_2-f_2\,\partial_if_1\right)\,.
\end{eqnarray} 
where $f(x)$ and ${\bf{X}}(x)$ are test functions and the PB of the
fundamental fields is defined in the following way,
\begin{equation}\label{funamenttal3}
\big\{\Pi_{ij}({\bf{x}}),q^{kl}({\bf{y}})\big\}=\frac{1}{2}\,\left(\delta^k_i\delta^l_j+\delta^k_j\delta^l_i\right)\delta({\bf{x}}-{\bf{y}})\,.
\end{equation} 
$C({\bf{X}})$ is the generator of the three-dimensional coordinate transformations, and
$C_0(f)$ corresponds to the transformation of the first and second
quadratic forms of the surface when it is deformed \cite{5Faddeev1982}.
Using the convention of eq. (\ref{funamenttal3}), the PBs of
eqs. (\ref{3com-1})-(\ref{3com-3}) can alternatively be written in the form
\begin{eqnarray}\label{3-comm-1}
X_1^i\big\{C_i,C_j\big\}X_2^j&=&\left(X_1^i\,X^j_{2,i}-X_2^i\,X^j_{1,i}\right)C_j\,,\\
\label{3-comm-2}
f\big\{C_0,C_i\big\}X^i&=&\left(fX^i_{,i}-X^if_{,i}\right)\,C_0\,,\\
\label{3-comm-3}
f_1\big\{C_0\,,C_0\big\}f_2&=&q^{ij}\left(f_1f_{2,i}-f_2f_{1,i}\right)C_j\,,
\end{eqnarray} 
which makes it easier comparing the PBs of the constraints derived in
\cite{5Faddeev1975,5Faddeev1982}, with the  
PBs of the tertiary first class constraints derived below.

The Hamiltonian of the first order EH action in terms of
$q^{ij}$ and $\Pi_{ij}$ as canonical variables was 
formulated in \cite{5Faddeev1975,5Faddeev1982},
where the authors use the metric-connection formulation of the first order EH action
as the basis of their analysis. The same Hamiltonian has
been independently formulated in \cite{5Schwinger1963}
where the starting point is the first order EH action in terms of
the vierbein $e^a_\mu$ and the connection $\omega^a_{\mu\nu}$. As in
\cite{5Faddeev1975,5Faddeev1982}, equations of motion are solved in
\cite{5Schwinger1963} in order to eliminate fields from the action,
compatible with the method of Faddeev and Jackiw \cite{5Faddeev1988}.

A novel canonical formulation of the
metric-connection formulation of the EH action in first order form
using the Dirac constraint formalism
\cite{5Dirac1950,5Dirac,5Gitman,5Hanson,5Henneaux1992,5Sudarshan,5Sundermeyer}
has been recently performed \cite{5Ghalati2007-2}. In this approach,
only equations of motion which correspond to second class
constraints are solved to eliminate fundamental fields from the
action, and the algebraic equations of motion which correspond to
first class constraints are used to generate constraints of higher order. The final form 
of the Hamiltonian action principle involves the fields $h$,
$h^i$, $H^{ij}$, $\bar t$ and $\bar \xi^i$ and their conjugate momenta
$\omega$, $\omega_i$, $\omega_{ij}$, $\Omega$ and $\Omega_i$. The
fields $h$, $h^i$ and $H^{ij}$ are   
  $h=\sqrt{-\mathfrak{g}}\,g^{00}$, $h^i=\sqrt{-\mathfrak{g}}\,g^{0i}$
  and  $H^{ij}=\frac{h^ih^j}{h}-h^{ij}$, where
  $h^{ij}=\sqrt{-\mathfrak{g}}\,g^{ij}$ and
  $\mathfrak{g}=\det(g_{\mu\nu})$. The momenta $\omega_{ij}$ are
  given\footnote{ For the definition of the 
  rest of the fields in terms of the metric
  $g^{\mu\nu}$ and the affine connection $\Gamma^\lambda_{\mu\nu}$ see
\cite{5Ghalati2007-2}.} as $\omega_{ij}=\Gamma^0_{ij}$. In terms
of these fields, the Hamiltonian action principle reads
as
\begin{eqnarray}\label{action}
S=\int dx \,\Big[\,\omega \,\dot h\!\!&+&\!\!\omega_i\, \dot
h^i+\omega_{ij}\,\dot H^{ij}+\Omega \,\dot{ 
\bar t}+\Omega_i\, \dot{{\bar \xi}}^i \\
\nonumber\!\!&-&\!\!\mathcal{H}^0_c-u\,\Omega-u^i\Omega_i-v\, 
\chi-v^i\chi_i-w \,\tau-w^i\tau_i\Big] 
\end{eqnarray}
where
\begin{eqnarray}\label{axadded}
\mathcal{H}^0_c&=&h\omega^2+h^i\omega \omega_i-\frac{d-3}{4(d-2)}H^{ij}\omega_i \omega_j-2\frac{h^m}{h}\,H^{ij}\omega_{im}\omega_j-\frac{1}{h}\,H^{ik}H^{jl}\omega_{jk}\omega_{il}\\\nonumber
&+&\frac{1}{h}h^i_{,\,j}h^j\omega_i+\frac{2}{h}\,h^i_{,\,j}H^{jk}\omega_{ik}-\frac{h^i}{h}\,H^{jk}_{\,\,,\,i}\omega_{jk}+\frac{1}{2(d-2)}H_{jk}H^{jk}_{\,\,,\,i}H^{im}\omega_m\\\nonumber
&-&\frac{1}{h}h^i_{,\,j}h^j_{,\,i}+\frac{1}{2}\,H^{jk}_{\,\,,\,i}H_{jq}H^{iq}_{\,,\,k}+\frac{1}{4}\,H^{ip}H_{kr,i}H^{kr}_{\,,\,p}+\frac{1}{4(d-2)}\,H^{ip}H_{jk}H^{jk}_{\,\,,\,i}H_{qr}H^{qr}_{\,\,,\,p}\\
\nonumber &+& \frac{1}{d-1}\,\frac{1}{h}\,\left(\chi^2-(2h\omega+h^i\omega_i)\chi\right)-\bar
\xi^i\chi_i-\frac{\bar t}{d-1}\,\chi+B^i\Lambda_i+B^{ij}\Lambda_i\Lambda_j\,,
\end{eqnarray}
where $u$, $u^i$, $v$, $v^i$, $w$ and $w^i$ are Lagrange multiplier
fields; $B^i$ and $B^{ij}$ are quantities that depend on the canonical
variables $h$, $h^i$, $H^{ij}$ and their conjugate momenta $\omega$,
$\omega_i$ and $\omega_{ij}$; $\Omega$ and $\Omega_i$ are primary
first class constraints; $\chi\approx0$ and $\chi_i\approx0$ are secondary first class
constraints of secondary stage
\begin{eqnarray}\label{ax22}
\chi&=&h^j_{,j}+h\,\omega-H^{jk}\,\omega_{jk}\,,\\ \label{ax23}
\chi_i&=&h_{,i}-h\,\omega_i\,;
\end{eqnarray}   
and $\tau\approx0$ and $\tau_i\approx0$ are secondary first class constraints of
tertiary stage,
\begin{eqnarray}\label{ax55}
\tau &=& -H^{ij}_{,ij}-(H^{ij}\omega_j)_{,i}-\frac{d-3}{4(d-2)}\,H^{ij}\omega_i\omega_j+\frac{1}{2(d-2)}\,H_{kl}H^{kl}_{\,\,,i}H^{ij}\omega_j\\ \nonumber
&-&\frac{1}{h}H^{ik}H^{jl}(\,\omega_{jk}\,\omega_{il}-\omega_{ik}\,\omega_{jl})+\frac{1}{2}H^{jk}_{\,\,,i}H_{jl}H^{il}_{,k} 
+\frac{1}{4}H^{ij}H_{kl,i}H^{kl}_{\,\,,j}\\ \nonumber
&+&
\frac{1}{4(d-2)}H^{ij}H_{kl}H^{kl}_{\,\,,i}H_{mn}H^{mn}_{\,\,,j}\,,
\end{eqnarray}
and
\begin{eqnarray}\label{ax51}
\tau_i&=&h\left(\frac{1}{h}\,H^{pq}\omega_{pq}\right)_{,i}+H^{pq}\omega_{pq,i}-2\left(H^{pq}\omega_{qi}\right)_{,p}\,.
\end{eqnarray} 
The constraints $\chi$, $\chi_i$, $\tau$ and $\tau_i$ are first
class and satisfy an unusual PB algebra as follows. For the PB of $\chi$ and $\chi_i$ we have
\begin{equation}\label{ax24}
\big\{\chi_i\,,\chi\big\} = \chi_i\,,
\end{equation}
while
\begin{equation}\label{ax25}
\big\{\chi_i\,,\chi_j\big\}=0=\big\{\chi\,,\chi\big\}\,.
\end{equation}
Also,
\begin{equation}\label{ax56}
\big\{\chi,\tau_i\big\}=0\,,
\end{equation}
\begin{equation}\label{ax57}
\big\{\chi,\tau\big\}=\tau\,,
\end{equation}
\begin{equation}\label{ax58}
\big\{\chi_i,\tau\big\}=0\,,
\end{equation}
and 
\begin{equation}\label{ax59}
\big\{\chi_i,\tau_j\big\}=0\,.
\end{equation}
The PBs of the constraints $\tau_i$ and $\tau$ are
nonlocal\footnote{We use the short notation
  $f\left\{X,Y\right\}g \equiv \int \!\int dxdy \,f(x)\left\{X(x),Y(y)\right\}g(y)$.}, as 
\begin{eqnarray}\label{ax60}
f\big\{\tau_i,\tau_j\big\}g&=&g(\partial_jf)\tau_i-f(\partial_ig)\tau_j\,,
\end{eqnarray}
\begin{eqnarray}\label{ax66}
f\big\{\tau,\tau\big\}g&=&\left(g\partial_if-f\partial_ig\right)\frac{H^{ij}}{h^2}\left(h\tau_j-H^{mn}\omega_{mn}\chi_j+2H^{mn}\omega_{mj}\chi_n\right)\,.
\end{eqnarray}
\begin{eqnarray}\label{ax67}
f\big\{\tau_i,\tau\big\}g = \nonumber &g&\!\!\!\!\frac{(fh)_{,i}}{h}\,\tau-fg_{,i}\tau-\frac{d-3}{2(d-2)}fgH^{kl}
\omega_k\left(\frac{\chi_l}{h}\right)_{,i}+f_{,k}\,g_{,l}\,H^{kl}\left(\frac{\chi_i}{h}\right)\\ \nonumber
&-&\frac{d-3}{2(d-2)}gf_{,k}H^{kl}\,\omega_l\left(\frac{\chi_i}{h}\right)+\frac{1}{2(d-2)}g\,f_{,m}\,H^{mn}H_{kl}H^{kl}_{\,\,,n}\left(\frac{\chi_i}{h}\right)\\ \nonumber
&+& f g_{,k} H^{kl}\left(\frac{\chi_l}{h}\right)_{,i}+ \frac{1}{2(d-2)}fgH^{mn}H_{kl}H^{kl}_{\,\,,m}\left(\frac{\chi_n}{h}\right)_{,i}\,,
\end{eqnarray}
where $f$ and $g$ are test functions. It may also be shown that the
Hamiltonian of eq. (\ref{axadded}) can be expressed in terms of the first
class constraints $\Omega$, $\Omega_i$, $\chi$, $\chi_i$, $\tau$ and $\tau_i$,
\begin{eqnarray}\label{axadded2}
\mathcal{H}&=&\tau+\frac{h^i}{h}\,\tau_i+\frac{h^i}{h}\,\chi_{,i}-\frac{1}{h^2}\,h^jh^i_{,j}\,\chi_i+\frac{d-2}{d-1}\,\frac{1}{h}\,H^{kl}\omega_{kl}\,\chi-\frac{h^i}{h}\,\omega\,\chi_i\\
\nonumber
&+&\frac{2}{h^2}\,h^kH^{ij}\omega_{ik}\,\chi_j+\frac{d-2}{d-1}\,\omega\,\chi+\frac{1}{d-1}\,\frac{1}{h}\,h^i_{,i}\,\chi-\frac{1}{d-1}\,\frac{h^l}{h}\,\omega_l\,\chi\\
\nonumber &-&\frac{t}{d-1}\,\chi-\bar\xi^i\,\chi_i+B^i\Lambda_i+B^{ij}\Lambda_i\Lambda_j\,.
\end{eqnarray} 

The (secondary) constraints $\chi$ and $\chi_i$, which have no counterpart in the
ADM Hamiltonian formulation of the first order EH action, are seen to arise because of the
consistency condition of vanishing of the primary constraints $\Omega$ and
$\Omega_i$, which are the momenta conjugate to the fields $\bar t$ and
$\bar \xi^i$,  which are in turn related to the connections
$\Gamma^\lambda_{\mu\nu}$ \cite{5Ghalati2007-2}.  

In the following section, we will show how the constraints $\chi$, $\chi_i$,
$\tau$ and $\tau_i$ of
eqs. (\ref{ax22}), (\ref{ax23}), (\ref{ax55}) and (\ref{ax51}) take a specially
simple form when the coordinates $H^{ij}$ are transformed to any set of
coordinates that depend \emph{only} on the metric $\gamma_{ij}$ of the
space-like surfaces $t=cons$. Two of the best 
sets of coordinates that can be used to replace the fields $H^{ij}$ are
the coordinates $q^{ij}$ used by Faddeev 
\cite{5Faddeev1975,5Faddeev1982} and the coordinates $\gamma_{ij}$ used by ADM
\cite{5Arnowitt1959-1,5Arnowitt1959-2,5Arnowitt1960,5Arnowitt1960-2,5Arnowitt}.
(These fields have been discussed in the previous 
chapter). In contrast to the variables $H^{ij}$
introduced in the previous chapter, the Faddeev variables
$q^{ij}=h^{0i}h^{0j}-h^{00}h^{ij}$ (where
$h^{\mu\nu}=\sqrt{-\mathfrak{g}}\,g^{\mu\nu}$) depend \emph{only} on the
components of the metric $\gamma_{ij}$ of the spacelike surfaces, since
\begin{eqnarray}\label{5Fadvar}
q^{ij}\gamma_{jk}&=&\gamma\,\delta^i_k\,,
\end{eqnarray} 
where $\gamma=\det(\gamma_{ij})$ \cite{5Faddeev1975,5Faddeev1982}. As
it will be seen in later sections, this has important
simplifying implications on the form of the algebra of the PB of
constraints and their dependence on the fundamental fields.
\section{Transforming to Faddeev variables}\label{II}
In the ADM Hamiltonian formulation of Faddeev
\cite{5Faddeev1975,5Faddeev1982}, the canonical coordinates
$q^{ij}=h^{0i}h^{0j}-h^{00}h^{ij}$ and their conjugate momenta
$\Pi_{ij}$ are the dynamical variables in the ``Hamiltonian'' and ``momentum''
constraints $C_0$ and $C_i$ of eqs. (\ref{mnm1}) and
(\ref{mnm2}), and thus the only dynamical variables in the
Hamiltonian formulation, subject to the constraints
$C_0\approx0$ and $C_i\approx0$. The fields
$\lambda=1+1/h^{00}$ and $\lambda^i=h^{0i}/h^{00}$ are non-dynamical
and act as Lagrange multiplier fields. In transition
from the variables $H^{ij}$ employed in the Dirac
Hamiltonian action principle of the first order EH action of
eq. (\ref{action}) to the Faddeev variables $q^{ij}$,\footnote{We will
  not transform the fields $h$ and $h^i$ to $\lambda$ and $\lambda^i$
  in the following, and will only consider transformation of the
  fields $q^{ij}$.} 
\begin{eqnarray}\label{5q}
q^{ij}&=&h\,H^{ij}\,,
\end{eqnarray}       
one must be careful that the momenta $\omega$, $\omega_i$ and $\omega_{ij}$ must
be transformed in such a way that \cite{5Gitman,5Goldstein,5Lanczos}
\begin{eqnarray}\label{5cantrans}
\omega\,\delta h+\omega_i\,\delta h^i+\omega_{ij}\,\delta H^{ij}&=&
\Pi \, \delta h+\Pi_i\,\delta h^i+\Pi_{ij}\,\delta q^{ij}\,,
\end{eqnarray} 
in order for the transformation to be canonical. This ensures
preservation of the properties of canonical invariants and the canonical equations of
motion. Eq. (\ref{5cantrans}) in turn results in the transformations 
\begin{equation}\label{5cantran2}
\omega=\Pi+\frac{1}{h}\,q^{ij}\,\Pi_{ij}\,\,\,\,\,\,\,\,\,\,\,\,\,\,\,\,\,\,\,\,\,\,\,\omega_i=\Pi_i\,\,\,\,\,\,\,\,\,\,\,\,\,\,\,\,\,\,\,\,\,\,\omega_{ij}=h\,\Pi_{ij} 
\end{equation} 
for the momenta. From eq. (\ref{5cantran2}), one observes that since
$\omega_{ij}=\Gamma^0_{ij}$,  the momenta $\Pi_{ij}$ agree with their definition in
\cite{5Faddeev1975,5Faddeev1982}, i.e. $\Pi_{ij}=\Gamma^0_{ij}/h$. We
note that the momentum corresponding to $h^i$  
remains unchanged as the transformation of eq. (\ref{5q}) does not
involve $h^i$. (This is also why the momenta $\Omega$ and $\Omega_i$ and their corresponding
canonical coordinates $\bar t$ and $\bar \xi^i$ do not appear in eq. (\ref{5cantrans})). In terms
of the new variables, the secondary first class constraints $\chi$ and
$\chi_i$ of eqs. (\ref{ax22}) and 
(\ref{ax23}) remarkably transform into 
\begin{eqnarray}\label{5ax22}
\tilde \chi&=&h^l_{,\,l}+h\,\Pi\\ \label{5ax23}
\tilde \chi_i&=&h_{,\,i}-h\,\Pi_i
\end{eqnarray}
respectively, while the tertiary first class constraint $\tau_i$ of
eq. (\ref{ax51}) transforms into 
\begin{eqnarray}\label{5ax51}
\tilde \tau_i=\left(q^{mn}\Pi_{mn}\right)_{,i}+q^{mn}\Pi_{mn,i}-2\left(q^{mn}\Pi_{mi}\right)_{,n}\,.
\end{eqnarray}
Surprisingly, the tertiary first class constraint
$\tau$ of eq. (\ref{ax55}) splits into several terms, some of which depend
on the secondary constraint $\tilde \chi_i$, 
\begin{eqnarray}\label{5ax55}
\tau&=&\frac{1}{h}\,\tilde
\tau-\frac{1}{2(d-2)}\,\frac{1}{h^2}\,q_{kl}\,q^{kl}_{\,\,,i}\,q^{ij}\tilde
\chi_j-\frac{d-3}{4(d-2)}\,\frac{1}{h^2}\,q^{ij}\tilde \chi_i \tilde
\chi_j+\frac{1}{h^3}\,h_{,j}\,q^{ij}\tilde \chi_i\\ \nonumber &+&
\left(\frac{1}{h^2}\,q^{ij}\tilde \chi_j\right)_{\!,\,i}\,. 
\end{eqnarray}
In eq. (\ref{5ax55}) we have 
\begin{eqnarray}\label{5ax55-2}
\tilde
\tau&=&-\,q^{ik}q^{jl}\left(\Pi_{jk}\Pi_{il}-\Pi_{ik}\Pi_{jl}\right)-q^{ij}_{\,\,,ij}+\frac{1}{2}\,q^{jk}_{\,\,,i}\,q_{jl}\,q^{il}_{\,\,,k}+\frac{1}{4}\, 
q^{ij}\,q_{kl,i}\,q^{kl}_{\,\,,j}\\ \nonumber
&+&\frac{1}{4(d-2)}\,q^{ij}\,q_{kl}\,q^{kl}_{\,\,,i}\,q_{mn}\,q^{mn}_{\,\,\,\,,j}\,. 
\end{eqnarray}
According to eq. (\ref{5ax55}), we may take the
constraint $\tilde \tau$ of eq. (\ref{5ax55-2}) to be the tertiary constraint arising from
the consistency condition that the time change of the constraint
$\tilde \chi$ must weakly vanish. The constraints $\tilde \tau_i$ and
$\tilde \tau$ of eqs. (\ref{5ax51},\ref{5ax55-2}) are 
indeed the constraints $C_i$ and $C_0$ of
eqs. (\ref{mnm1},\ref{mnm2}) in the Faddeev
Hamiltonian formulation of the first order EH action. 

It is seen from eqs. (\ref{5ax22}-\ref{5ax51},\ref{5ax55-2}) that,
when written in terms of the variables $h$, 
$h^i$, $q^{ij}$ and their conjugate momenta $\Pi$, $\Pi_i$ and $\Pi_{ij}$, the constraints
$\tilde \chi$ and $\tilde \chi_i$ depend only on the variables $h$,
$h^i$ and their conjugate momenta $\Pi$ and $\Pi_i$, while the
constraints $\tilde \tau_i$ and $\tilde \tau$ depend exclusively on the
canonical variables $q^{ij}$ and their conjugate momenta
$\Pi_{ij}$. Thus, the variables $h$ and $h^i$ and their conjugate
momenta $\Pi$ and $\Pi_i$ are seen to decouple from the variables $q^{ij}$
and their conjugate momenta $\Pi_{ij}$ in formation of the first class constraints. 

Since the PB (as well as the DB, because it is defined in terms of the PB)
is invariant under canonical transformations, we see that under the transformations of
eqs. (\ref{5q}) and (\ref{5cantran2}), the PBs of eqs. (\ref{ax24}) and
(\ref{ax25}) imply that
\begin{eqnarray}\label{5ax24}
\big\{\tilde \chi_i\,,\tilde \chi\big\} &=& \tilde \chi_i\,,\\ \label{5ax25}
\big\{\tilde \chi_i\,,\tilde \chi_j\big\}&=&0\,,\\ \label{5ax26}
\big\{\tilde \chi\,,\tilde \chi\big\} &=&0\,.
\end{eqnarray}  
There is a remarkable way of obtaining the algebra of the PBs of
the new constraints $\tilde \tau_i$ and $\tilde \tau$ of eqs. (\ref{5ax51})
and (\ref{5ax55}) directly from the PBs of
eqs. (\ref{ax60},\ref{ax66},\ref{ax67}) of the constraints $\tau_i$
and $\tau$. We note that the 
constraint $\tilde \tau_i$ of eq. (\ref{5ax51}) can be obtained from
the constraint $\tau_i$ of eq. (\ref{ax51}) by substituting
$h=1$.\footnote{There must also be an appropriate identification of the
  corresponding fields and momenta; i.e, by replacing $H^{ij}$ and
  $\omega_{ij}$ with $q^{ij}$ and $\Pi_{ij}$ in the expression
  obtained.} Since
$\tau_i$ does not depend on the momentum $\omega$ conjugate to $h$, the
latter is passive in computing the PB of eq. (\ref{ax60}), that is, since
$\tau_i$ is independent of $\omega$, it makes
no difference if we were to set $h=1$ before or after the PB
$\big\{\tau_i,\tau_j\big\}$ is computed. Therefore, 
we may set $h=1$ in both sides of eq. (\ref{ax60}) and conclude that
\begin{eqnarray}\label{5ax60}
f\big\{\tilde \tau_i,\tilde \tau_j\big\}g=gf_{,j}\tilde
\tau_i-fg_{,i}\tilde \tau_j\,.
\end{eqnarray}   
since $\tilde \tau_i$ depends on $q^{ij}$ and $\Pi_{ij}$ in the same way
that $\tau_i$ depends on $H^{ij}$ and $\omega_{ij}$ once we set $h=1$
in $\tau_i$. 

In a similar way, we may compute the PBs $\big\{\tilde \tau,\tilde
\tau\big\}$ and $\big\{\tilde \tau_i,\tilde \tau \big\}$ from the PBs
$\big\{\tau,\tau\big\}$ and $\big\{\tau_i,\tau\big\}$ of eqs. (\ref{ax66}) and (\ref{ax67})
without explicitly computing these PBs using the fundamental PBs among
the new canonical variables. The constraint 
$\tau$ of eq. (\ref{ax55}) reduces to $\tilde \tau$ of
eq. (\ref{5ax55}) by substituting $h=1$ and $\omega_i=0$ in
eq. (\ref{ax55}). Since $\tau$ has no dependence on either the momenta $\omega$
conjugate to $h$ or the field $h^i$ conjugate to the momenta
$\omega_i$, one may set $h=1$ and $\omega_i=0$ either before or after
the PB $\big\{\tau,\tau\big\}$ of eq. (\ref{ax66}) is computed, and obtain
the same quantity. This implies that
\begin{eqnarray}\label{5ax66}
f\big\{\tilde \tau,\tilde
\tau\big\}g=\left(gf_{,i}-fg_{,i}\right)q^{ij}\,\tilde \tau_j\,.
\end{eqnarray}
In much the same way, one may set $h=1$ and $\omega_i=0$ in both sides
of eq. (\ref{ax67}), and conclude that 
\begin{eqnarray}\label{5ax67}
f\big\{\tilde \tau_i,\tilde \tau\big\}g &=& \left(gf_{,i}-fg_{,i}\right)\tilde\tau\,.
\end{eqnarray}
The PBs of the first class constraints $\tilde \tau_i$ and $\tilde
\tau$ of eqs. (\ref{5ax60}), (\ref{5ax66}) and (\ref{5ax67}) are
indeed identical to the PBs of eqs. (\ref{3-comm-1}), (\ref{3-comm-3})
and (\ref{3-comm-2}) of the ADM Hamiltonian formulation of 
Faddeev if we identify $\tilde \tau_i$ and $\tilde \tau$ of eqs. (\ref{5ax51})
and (\ref{5ax55-2}) with the constraints $C_i$ and $C_0$ of
eqs. (\ref{mnm1}) and (\ref{mnm2}) derived by Faddeev, considering
that in eq.\ (3.13) of \cite{5Faddeev1982} the fundamental PBs are defined as
\begin{equation}\label{funamenttal}
\big\{\Pi_{ij}({\bf{x}}),q^{kl}({\bf{y}})\big\}=\frac{1}{2}\,\left(\delta^k_i\delta^l_j+\delta^k_j\delta^l_i\right)\delta({\bf{x}}-{\bf{y}})\,.
\end{equation} 
In fact, an explicit calculation of the expressions of
eqs. (\ref{mnm1},\ref{mnm2}) using eqs. (\ref{covariantderiv0}-\ref{covariantderiv2}) shows that 
\begin{eqnarray}\label{mhat1}
C_k&=&(q^{il}\Pi_{il})_{,\,k}+q^{il}\Pi_{il,k}-2\,(q^{il}\Pi_{ik})_{,\,l}\\\label{mhat2}
C_0&=&\,q^{ik}q^{mn}\left(\Pi_{ik}\Pi_{mn}-\Pi_{im}\Pi_{kn}\right)-q^{ij}_{\,\,,ij}+\frac{1}{2}\,q^{jk}_{\,\,,i}\,q_{jl}\,q^{il}_{\,\,,k}\,,\\
\nonumber &+&\frac{1}{4}\, 
q^{ij}\,q_{kl,i}\,q^{kl}_{\,\,,j}+\frac{1}{8}\,q^{ij}\,q_{kl}\,q^{kl}_{\,\,,i}\,q_{mn}\,q^{mn}_{\,\,\,\,,j}\,,
\end{eqnarray}
which are the constraints $\tilde \tau_i$ and $\tilde \tau$ of
eqs. (\ref{5ax51},\ref{5ax55-2}) when $d=4$.

We now express the Hamiltonian of eq. (\ref{axadded2}) in terms of the
new variables $h$, $h^i$, $q^{ij}$ and their 
conjugate momenta. Under the transformations of eqs. (\ref{5q}) and (\ref{5cantran2}), one
obtains
\begin{eqnarray}\label{5Hamiltoni}
\mathcal{H}_c&=&\frac{1}{h}
\,\tilde \tau+\frac{h^i}{h}\,\tilde
\tau_i-\frac{1}{2(d-2)}\,\frac{1}{h^2}\,q_{kl}q^{kl}_{\,\,,i}\,q^{ij}\tilde
\chi_j-\frac{d-3}{4(d-2)}\,\frac{1}{h^2}\,q^{ij}\tilde \chi_i\tilde
\chi_j\\ \nonumber &+&\frac{h_{,j}}{h^3}\,q^{ij}\chi_i-\frac{h^j}{h^2}\,h^i_{,j}\tilde
\chi_i+2\,\frac{h^m}{h^2}\,q^{il}\Pi_{lm}\tilde
\chi_i+\frac{h^i}{h^2}\,h_{,i}\,\tilde
\chi-\frac{h^i}{h}\,\Pi\,\tilde
\chi_i\\ \nonumber &-&\frac{h^i}{h^2}\,q^{mn}\Pi_{mn}\,\tilde
\chi_i-\frac{1}{d-1}\,\frac{h^l}{h}\,\Pi_l\,\tilde
\chi+\frac{d-2}{d-1}\frac{1}{h}\left(h\Pi \,\tilde 
\chi+2\,q^{ij}\Pi_{ij} \,\tilde
\chi-h^l_{,l}\,\tilde\chi\right) \\ \nonumber &-& 
\frac{\bar t}{d-1}\,\tilde \chi-\bar \xi^i\tilde \chi_i+\tilde B^i\,\tilde \Lambda_i+\tilde B^{ij}\, \tilde \Lambda_{i} \tilde \Lambda_j\,,
\end{eqnarray}
after a surface term has been dropped. 
The Hamiltonian of eq. (\ref{5Hamiltoni}) contains the ``Hamiltonian'' and ``momentum''
constraints $\tilde \tau=C_0$ and $\tilde 
\tau_i=C_i$ appearing in the Hamiltonian of the action of eq. (\ref{5Faddeev}) derived
by Faddeev, but in addition it incorporates terms proportional to the
secondary first class constraints $\tilde \chi$ and $\tilde \chi_i$,
as well as the terms proportional to the primary first class
constraints $\Omega$ and $\Omega_i$, which are present in $\tilde \Lambda_i$.
 
The Hamiltonian action principle for the Hamiltonian of eq. (\ref{5Hamiltoni}),
therefore, takes the form
\begin{eqnarray}\label{5action}
S=\int dx \,\Big[\,\Pi \,\dot h\!\!&+&\!\!\Pi_i\, \dot
h^i+\Pi_{ij}\,\dot q^{ij}+\Omega \,\dot {\bar 
t}+\Omega_i\, {\dot {\bar \xi}}^i \\
\nonumber\!\!\!\!&-&\!\!\mathcal{H}_c-u\,\Omega-u^i\,\Omega_i-v\,\tilde 
\chi-v^i\tilde \chi_i-w \,\tilde \tau-w^i\tilde \tau_i\Big] 
\end{eqnarray} 
where $\mathcal{H}_c$ is given by eq. (\ref{5Hamiltoni}) and $\tilde
\chi$, $\tilde \chi_i$, $\tilde \tau_i$ and $\tilde \tau$ are given by
eqs. (\ref{5ax22}), (\ref{5ax23}), (\ref{5ax51}) and  (\ref{5ax55}).
In contrast with the Faddeev action of eq. (\ref{5Faddeev}), we see that in
the action of eq. (\ref{5action}), besides the fields $q^{ij}$ and
$\Pi_{ij}$, the fields $h$, $h^i$, $\bar t$, $\bar \xi^i$ and their
corresponding momenta $\Pi$, $\Pi_i$, $\Omega$ and $\Omega_i$ appear
to be dynamical. However, these fields are subject to more 
constraints, namely, $\Omega\approx0$, $\Omega_i\approx0$, $\tilde
\chi \approx 0$, $\tilde \chi_i\approx 0$, $\tilde \tau_i \approx 0$
and $\tilde \tau\approx 0$, so that the number of
degrees of freedom turns out to be counted the same as that of the
ADM. 

One may apply the reduction method of Faddeev and Jackiw \cite{5Faddeev1988} to the action of
eq. (\ref{5action}), in which one considers all the canonical
variables and Lagrange multipliers in the action at the same footing as
\emph{fields}. Equations of motion 
for the fields $u$, $u^i$, $v$ and $v^i$ result in $\Omega=0$,
$\Omega_i=0$, $\tilde \chi=0$ and $\tilde \chi_i=0$. These equations may be solved for the
fields $\Omega$, $\Omega_i$, $\Pi$ and $\Pi_i$. Upon substituting these
solutions into the action of 
eq. (\ref{5action}) all terms coming from the Hamiltonian of eq. (\ref{5Hamiltoni}) 
vanish except for the terms proportional to $\tilde \tau$ and $\tilde
\tau_i$, and the kinetic term becomes
\begin{eqnarray}\label{5surface}
\Pi \,\dot h+\Pi_i \,\dot h^i+\Pi_{ij}\,\dot q^{ij}&=&
\left(\frac{h_{,\,l}\,h^l}{h}\right)_{\!,\,0}-\left(\frac{\dot h\,
h^l}{h}\right)_{\!,\,l}+\Pi_{ij}\,\dot q^{ij}\,.
\end{eqnarray}  
The first two terms on the right hand side being total derivatives
may be dropped from the action of eq. (\ref{5action}), which would now take the form 
\begin{eqnarray}\label{5action2}
S=\int dx \,\Big[\,\Pi_{ij}\,\dot q^{ij}-\frac{1}{h}\,\tilde
\tau-\frac{h^i}{h}\,\tilde \tau_i-w \,\tilde \tau-w^i\tilde \tau_i\Big]\,, 
\end{eqnarray}
which is the Faddeev version of the ADM action of eq. (\ref{5Faddeev}). 

In the context of the Dirac constraint formalism, however,
the first class constraints $\Omega$, $\Omega_i$, $\tilde \chi$ and
$\tilde \chi_i$ may be solved only if appropriate 
gauge fixing conditions for all first class constraints are assumed
\cite{5Henneaux1992,5Sundermeyer}. Together with the first class
constraints $\Omega$, $\Omega_i$, $\chi$ and $\chi_i$, their gauge
constraints may then be turned into strong equations while the PB is
replaced with the appropriate DB. These equations
may then be solved in order to eliminate fields from the action of
eq. (\ref{5action}).

The introduction of gauge fixing conditions for the action of
eq. (\ref{5action}), however, requires a knowledge of the gauge
transformations of this action beforehand \cite{5Sundermeyer}. To obtain the
generator of the gauge transformations all first class constraints
$\Omega$, $\Omega_i$, $\tilde \chi$, $\tilde \chi_i$, $\tilde \tau$
and $\tilde \tau_i$ are required
\cite{5Castellani1982-2,5Dirac,5Henneaux1992}. Once a set of
admissible gauge constraints are assumed and the first 
class constraints $\Omega$, $\Omega_i$, $\tilde \chi$ and $\tilde \chi_i$ are turned into
second class, they no longer act as generators of gauge
transformations. Therefore, gauge fixing of the action
of eq. (\ref{5action}) will result in losing some information about the
generator of the gauge transformations of this action. The situation is similar to
the gauge fixing of the
``algebraic'' constraint $e^0\approx0$ ($e^0$ is the momentum
conjugate to $A_0$, the temporal component of $A_\mu$) in the
Hamiltonian formulation of Maxwell gauge fields by using the gauge
constraint $A_0\approx0$, and subsequent loss of the
generator of the gauge transformation for $A_0$. (See
\cite{5Sundermeyer} for a discussion of the canonical formulation of
the Maxwell gauge fields.)  

\section{Transforming to ADM variables}\label{III}
In the original formulation of ADM, the EH action to start with is written in terms
of the covariant components of the metric $\gamma_{ij}$ of the
spacelike surfaces characterized by a time coordinate $t=$ cons., and
the components $N$ and $N_i$ of the lapse and shift functions
defined in terms of the metric $g_{\mu\nu}$ of the four dimensional
embedding space in eq. (\ref{ghd6}). In the action of eq. (\ref{5action}), a transformation 
from the variables $q^{ij}$ to $\gamma_{ij}$ using eq. (\ref{5Fadvar}),
\begin{eqnarray}\label{5gamma}
q^{ij}&=&\gamma\,\gamma^{ij}\qquad\qquad\qquad\gamma=\det{\gamma_{ij}}\,, 
\end{eqnarray}
must be accompanied by appropriate transformations of
the momenta $\Pi_{ij}$ conjugate to $q^{ij}$ to the momenta $\pi^{ij}$
conjugate to $\gamma_{ij}$, so that 
\begin{eqnarray}\label{5trans22}
\Pi_{ij} \,\delta q^{ij}&=&\pi^{ij}\delta \gamma_{ij}\,.
\end{eqnarray}   
This implies that the momenta should be transformed in the following way,
\begin{eqnarray}\label{5cantran22}
\Pi_{ij}&=&-\,\gamma^{-1}\left(\gamma_{ia}\gamma_{jb}-\frac{1}{d-2}\,\gamma_{ij}\gamma_{ab}\right)\pi^{ab}\,.
\end{eqnarray} 
Once again, one may directly check that if $\Pi_{ij}=\Gamma^0_{ij}/h$,
  as defined in \cite{5Faddeev1975,5Faddeev1982} and eq. (\ref{5cantran2}), then the momenta
  $\pi^{ij}$ defined in eq. (\ref{5cantran22}) are the same as 
  the ADM momenta $\pi^{ij}$ given in eq. (\ref{ghd5}). Under the
  canonical transformations of eqs. (\ref{5gamma}) and 
(\ref{5cantran22}), the constraints $\tilde \chi$ and $\tilde \chi_i$
of eqs. (\ref{5ax22}) and (\ref{5ax23}) remain unchanged. The momentum
constraint $\tilde \tau_i$ of eq. (\ref{5ax51}), however, transforms to
\begin{eqnarray}\label{5taui}
\tilde
\tau_i&=&-\,\pi^{ab}\gamma_{ab,i}+2\left(\gamma_{ib}\pi^{ab}\right)_{,a}\\
\nonumber &=&-\mathcal{H}_i\,,
\end{eqnarray}
where the ADM momentum constraint $\mathcal{H}^i$ is given by
eq. (\ref{ghd8-2}). Also, from eq. (\ref{5ax55-2}) we find that
\begin{eqnarray}\label{5tau}
\tilde \tau&=&
-\left(\pi^{ij}\,\pi_{ij}-\frac{1}{d-2}\,\left(\pi^l_{\,l}\right)^2\right)+\gamma\,\mathcal{R}\\
 \nonumber &=&-\,\gamma^{1/2}\,\mathcal{H},
\end{eqnarray}
where the ADM Hamiltonian constraint $\mathcal{H}$ is given by
\begin{eqnarray}\label{AAAAAAAAAA55hjash}
\mathcal{H}&=&\gamma^{-1/2}\left(\pi^{ij}\pi_{ij}-\frac{1}{d-2}\,\left(\pi^l_{\,l}\right)^2\right)-\gamma^{1/2}\,\mathcal{R}\,.
\end{eqnarray}
Therefore, in terms of $h$, $h^i$ and $\gamma_{ij}$,
the action of eq. (\ref{5action}) becomes 
\begin{eqnarray}\label{5action3}
S=\int dx \,\Big[\,\Pi \,\dot h\!\!&+&\!\!\Pi_i\, \dot
h^i+\pi^{ij}\,\dot \gamma_{ij}+\Omega \,\dot 
{\bar t}+\Omega_i\, \dot {\bar \xi}^i \\
\nonumber\!\!&-&\!\!\mathcal{H}'_c-u\,\Omega-u^i\Omega_i-v\tilde 
\chi-v^i\tilde \chi_i-w \,\tilde \tau-w^i\tilde \tau_i\Big]\,,  
\end{eqnarray}
where $\mathcal{H}'_c$ is the Hamiltonian of eq. (\ref{5Hamiltoni})
transformed under eqs. (\ref{5gamma}) and (\ref{5cantran22}),
\begin{eqnarray}\label{5Hamiltoni5}
\mathcal{H}'_c&=&-\,\frac{1}{h}
\,\gamma^{1/2}\,\mathcal{H}-\frac{h^i}{h}\,
\mathcal{H}_i+\frac{1}{2\,h^2}\,\gamma\,\gamma_{kl}\gamma^{kl}_{\,\,,i}\,\gamma^{ij}\tilde
\chi_j-\frac{d-3}{4(d-2)}\,\frac{1}{h^2}\,\gamma\,\gamma^{ij}\tilde \chi_i\tilde
\chi_j\\ \nonumber &+&\frac{h_{,j}}{h^3}\,\gamma\,\gamma^{ij}\tilde \chi_i-\frac{h^j}{h^2}\,h^i_{,j}\,\tilde
\chi_i-2\,\frac{h^m}{h^2}\,\gamma_{mj}\pi^{ij}\,\tilde
\chi_i+\frac{1}{d-1}\,\frac{h^i}{h^2}\,\gamma_{ab}\,\pi^{ab}\,\tilde
\chi_i+\frac{h^i}{h^2}\,h_{,i}\,\tilde\chi\\ \nonumber &-&\frac{h^i}{h}\,\Pi\,\tilde
\chi_i-\frac{1}{d-1}\,\frac{h^l}{h}\,\Pi_l\,\tilde
\chi+\frac{d-2}{d-1}\frac{1}{h}\left(h\Pi \,\tilde 
\chi-h^l_{,l}\,\tilde\chi\right)+\frac{2}{d-1}\,\frac{1}{h}\,\gamma_{ab}\,\pi^{ab}\tilde
\chi\\ \nonumber &-& 
\frac{\bar t}{d-1}\,\tilde \chi-\bar \xi^i\tilde \chi_i+\tilde B^i\,\tilde \Lambda_i+\tilde B^{ij}\, \tilde \Lambda_{i} \tilde \Lambda_j\,.
\end{eqnarray}
We have thus achieved a Hamiltonian formulation of the EH action in
terms of the variables $h$, $h^i$, $\gamma_{ij}$ and their
corresponding momenta $\Pi$, $\Pi_i$ and $\pi^{ij}$. 
The ADM Hamiltonian constraint $\mathcal{H}$ appears with a coefficient $\gamma^{1/2}$. 
Such a factor can be combined with the field $h$ in
the action of eq. (\ref{5action3}) in order to introduce the lapse and shift
functions $N$ and $N^i$ and their conjugate momenta ``as canonical
variables''. In terms of the metric $g_{\mu\nu}$ the lapse
and shift functions $N$ and $N^i$ are defined 
as\footnote{We note that it \emph{makes difference} whether we use
$N^i$ or its ``covariant'' component $N_i=\gamma_{ij}\,N^{j}$ as the
canonical variable.} 
\begin{eqnarray}\label{5lapse}
g^{00}=-\,\frac{1}{N^2}\,,\,\,\,\,\,\,\,\,\,\,\,\,\,\,\,\,\,\,\,\,\,g^{0i}=\frac{N^i}{N^2}\,.
\end{eqnarray}
Consequently, in terms of the metric $\gamma_{ij}$ of the spacelike surfaces and the variables
$h=\sqrt{-\mathfrak{g}}\,g^{00}$ and
$h^i=\sqrt{-\mathfrak{g}}\,g^{0i}$ we have 
\begin{eqnarray}\label{5ttran}
h=-\,{\gamma}^{1/2}\,\frac{1}{N}\,,\,\,\,\,\,\,\,\,\,\,\,\,\,\,\,\,\,\,\,\,\,\,\,h^{0i}={\gamma}^{1/2}\,\frac{N^i}{N}\,.
\end{eqnarray}   
As eqs. (\ref{5ttran}) depend on the metric $\gamma_{ij}$, we must require that the momenta $\Pi$, $\Pi_i$  and $\pi^{ij}$
conjugate to $h$, $h^i$ and $\gamma_{ij}$ transform to the canonical momenta $p$,
$p_i$ and $p^{ij}$ conjugate to $N$ and $N^i$ and $\gamma_{ij}$ in
such a way that
\begin{eqnarray}\label{5oneform}
\Pi \,\delta h+\Pi_i\, \delta h^i+\pi^{ij}\,\delta
\gamma_{ij}&=&p\,\delta N+p_i\,\delta N^i+p^{ij}\delta \gamma_{ij}.
\end{eqnarray}
This implies that
\begin{eqnarray}\label{5tootra}
\Pi&=&\frac{1}{\sqrt{\gamma}}\,N\left(Np+N^ip_i\right)\,,\\ \label{5tootra2}
\Pi_i&=&\frac{1}{\sqrt{\gamma}}\,N\,p_i\,,\\ \label{5tootra3}
\pi^{ij}&=&p^{ij}+\frac{1}{2}\,\gamma^{ij}Np\,.
\end{eqnarray}
The momenta $p^{ij}$ defined in eq. (\ref{5tootra3}) are not the 
same as the ADM momenta $\pi^{ij}$ defined in eq. (\ref{ghd5}). Under
the canonical transformations of eqs. (\ref{5ttran}) and
(\ref{5tootra}-\ref{5tootra3}), the constraints $\tilde \chi_i$ and
$\tilde \chi$ transform to
\begin{eqnarray}\label{5totaltra}
\tilde \chi_i&=& -\left(\frac{\sqrt{\gamma}}{N}\right)_{\!,\,i}+p_i\,,\\
\label{5totaltra2}
\tilde \chi&=&
-\left(-\left(\frac{\sqrt{\gamma}}{N}\right)_{\!,\,i}+p_i\right)N^i+\left(\frac{\sqrt{\gamma}}{N^2}\,N^i_{,i}-p\right)N\,;
\end{eqnarray}
and for the constraints $\tilde \tau_i$ and $\tilde \tau$ one finds that
\begin{eqnarray}\label{5totaltra3}
\frac{h^i}{h}\,\tilde
\tau_i&=&N^i\tilde{\mathcal{H}_i}=N^i\left(\mathfrak{H}_i+\frac{1}{2}\,Np\,\gamma^{ab}\gamma_{ab,i}-\left(Np\right)_{,i}\right)\,,\\  
\label{5totaltra4}
\frac{1}{h}\,\tilde\tau&=&N\,\tilde{\mathcal{H}}=N\left(\mathfrak{H}\,-\frac{d-1}{4(d-2)}\,\frac{1}{\sqrt{\gamma}}\,(Np)^2-\frac{1}{d-2}\,\frac{N}{\sqrt{\gamma}}\,p\,\gamma_{ab}\,p^{ab}\right)\,,
\end{eqnarray}
where
\begin{eqnarray}\label{55hjash2342}
\mathfrak{H}_i&=&-\left(-p^{ab}\gamma_{ab,i}+2\left(\gamma_{ib}\,p^{ab}\right)_{\!,\,a}\right),\\ \label{55hjash2}
\mathfrak{H}&=&\gamma^{-1/2}\left(p^{ij}p_{ij}-\frac{1}{d-2}\,\left(p^l_{\,l}\right)^2\right)-\gamma^{1/2}\,\mathcal{R}\,.
\end{eqnarray}
The canonical transformations of the variables $h$ and $h^i$
to the variables $N$ and $N^i$ result in the dependence of the constraints $\tilde \chi_i$ and
$\tilde \chi$ on the metric $\gamma_{ij}$ of the spacelike surfaces,
and in the constraints $\tilde \tau_i$ and $\tilde \tau$ receiving contributions
from the fields $N$ and $N^i$ and their conjugate momenta $p$ 
and $p_i$\,. 

Once again, we may apply the method of Faddeev and Jackiw to the
action of eq. (\ref{5action3}) after the fields $h$ and $h^i$ are
canonically transformed to $N$ and $N^i$ according to
eqs. (\ref{5ttran},\ref{5tootra},\ref{5tootra2},\ref{5tootra3}). The
equations of motion of the fields $u$, $u^i$, $v$ and $v^i$ result in
$\Omega=0$, $\Omega_i=0$, $\tilde \chi=0$ and $\tilde \chi_i=0$, where
$\tilde \chi_i$ and $\tilde \chi$ are given by
eqs. (\ref{5totaltra},\ref{5totaltra2}). We may then solve these
constraints for $\Omega$, $\Omega_i$, $p$ and $p_i$
and insert their solutions in the action, and in particular in
eqs. (\ref{5totaltra3},\ref{5totaltra4}). The kinetic part of the
action then transforms to 
\begin{eqnarray}\label{55hjgjjt}
p\,\dot N+p_i \,\dot
N^i+p^{\,ij}\,\dot
\gamma_{ij}&=&\left(N^i\left(\frac{\gamma^{1/2}}{N}\right)_{\!,\,i}\right)_{\!,\,0}\!\!-\left(N^i\left(\frac{\gamma^{1/2}}{N}\right)_{\!,\,0}\right)_{\!,\,i}\\
\nonumber &+& p^{\,ij}\,\dot \gamma_{ij}+\frac{1}{N}\,N^i_{,i}\,\dot \gamma^{1/2}\,.
\end{eqnarray}   
The first two terms on the right hand side may be dropped from the action since they are
total derivatives. The appropriate Darboux transformation \cite{5Faddeev1988} associated
with the reduced kinetic term is 
\begin{eqnarray}\label{darboux}
\tilde p^{ij}&=&p^{ij}+\frac{\sqrt{\gamma}}{2N}\,N^l_{,l}\,\gamma^{ij}\,,
\end{eqnarray}
by which the kinetic term takes the standard form  
\begin{eqnarray}\label{55kdf}
p^{\,ij}\,\dot \gamma_{ij}+\frac{1}{N}\,N^i_{,i}\,\dot
\gamma^{1/2}&=&\tilde p^{ij}\,\dot \gamma_{ij}\,.
\end{eqnarray}
The momenta $\tilde p^{ij}$ defined in eq. (\ref{darboux}) are the
same as the ADM momenta defined in eqs. (\ref{ghd5},\ref{5cantran22}).
Upon transforming the action under the transformations of
eq. (\ref{darboux}), it is seen that $\tilde{\mathcal{H}}_i$ and
$\tilde{\mathcal{H}}$ of eqs. (\ref{5totaltra3},\ref{5totaltra4})
transform into the ADM momentum and Hamiltonian constraints
$\mathcal{H}_i$ and $\mathcal{H}$ of
eqs. (\ref{5taui}) and (\ref{AAAAAAAAAA55hjash}). The reduced action is
therefore
\begin{eqnarray}\label{admredu}
S&=&\int dx \,\Big[\,\tilde p^{ij}\,\dot \gamma_{ij}-N\,\tilde
{\mathcal{H}}-N^i\,\tilde {\mathcal{H}}_i-w \,\tilde \tau-w^i\tilde \tau_i\Big]\,, 
\end{eqnarray} 
which is the ADM action upon a redefinition of the
Lagrange multipliers $w$ and $w_i$.
  
Instead of introducing the lapse and shift functions $N$ and $N^i$ in
the action of eq. (\ref{5action3}), one may choose
the most natural choice of coordinates that avoid mixing of the
canonical fields in formation of the constraints, i.e. the ``densitized'' lapse function 
\begin{eqnarray}\label{5dlapse}
\alpha=N\gamma^{-1/2}
\end{eqnarray}
and the shift functions $\alpha^i$, which are defined as in the ADM
approach.\footnote{We note that $\alpha=\lambda^0-1$ and
$\alpha^i=\lambda^i$, where $\lambda^0$ and $\lambda^i$ are the Lagrange
multipliers appearing in eq. (\ref{5Faddeev}).} From eq. (\ref{5ttran}) one then has, 
\begin{eqnarray}\label{5htoalpha}
h=-\frac{1}{\alpha}\,,\,\,\,\,\,\,\,\,\,\,\,\,\,\,\,\,\,\,\,\,\,\,\,\,\,\,\,\,\,\,h^i=\frac{\alpha^i}{\alpha}\,
\end{eqnarray} 
and consequently
\begin{eqnarray}\label{5dlapsemomenta}
\Pi=\alpha\left(\alpha\,
\pi+\alpha^i\pi_i\right)\,,\,\,\,\,\,\,\,\,\,\,\,\,\,\,\,\,\,\,\,\,\,\,\,\,\,\Pi_i=\alpha\,
\pi_i\,,
\end{eqnarray}
where $\pi$ and $\pi_i$ are the momenta conjugate to $\alpha$ and
$\alpha_i$. We see that, in contrast with eqs. (\ref{5tootra})-(\ref{5tootra3}), the fields
$\gamma_{ij}$ and their conjugate momenta do not enter the
transformations of eqs. (\ref{5dlapsemomenta})\,. The constraints
$\tilde \chi$ and $\tilde \chi_i$ of eqs. (\ref{5ax22}) and
(\ref{5ax23}) then transform into
\begin{eqnarray}\label{5ax22adm}
\tilde \chi &=&
\alpha\left(\frac{\alpha^i_{,i}}{\alpha^2}-\pi\right)-\alpha^i\left(-\left(\frac{1}{\alpha}\right)_{,i}+\pi_i\right)\,,\\
\label{5ax23adm}
\tilde \chi_i&=&-\left(\frac{1}{\alpha}\right)_{,i}+\pi_i\,,
\end{eqnarray}
which depend only on a subset of the canonical variables;
$\alpha$, $\alpha^i$ and their conjugate momenta $\pi$ and $\pi_i$.\footnote{
We note that at this stage the constraints $\tilde \chi\approx 0$ and $\tilde
\chi_i \approx 0$ of eqs. (\ref{5ax22adm}) and (\ref{5ax23adm}) might be
replaced with the constraints $\phi_i\approx 0$ and $\phi \approx 0$
through $\tilde \chi_i=-\,\phi_i$ and $\tilde
\chi=\alpha\,\phi+\alpha^i\phi_i$, where
$\phi_i=-\frac{\alpha_{,i}}{\alpha^2}-\pi_i$ and $\phi=\frac{\alpha^i_{,i}}{\alpha^2}-\pi$, 
however, since such an identification does not show to be particularly
illuminating, we won't pursue it at this stage.} The constraints $\tilde
\tau$ and $\tilde \tau_i$ are seen to depend only on the rest of the
canonical coordinates, i.e. $\gamma_{ij}$ and their conjugate momenta
$\pi^{ij}$, and they remain intact under the transformations of eqs. (\ref{5htoalpha}) and
(\ref{5dlapsemomenta}). We thus introduce the quantities  
\begin{eqnarray}\label{5axtrauma}
\bar{\mathcal{H}}&=&-\,\tilde
\tau=\left(\pi^{ij}\,\pi_{ij}-\frac{1}{d-2}\,\left(\pi^l_{\,l}\right)^2\right)-\gamma\,\mathcal{R}\,,\\
\label{5axtrauma2}
\mathcal{H}_i&=&-\tilde \tau_i=\pi^{ab}\gamma_{ab,i}-2\left(\gamma_{ib}\pi^{ab}\right)_{,a}\,,
\end{eqnarray}
and for the action of eq. (\ref{5action3}) we obtain
\begin{eqnarray}\label{5action32}
S=\int dx \,\Big[\,\pi \,\dot \alpha\!\!&+&\!\!\pi_i\, \dot
\alpha^i+\pi^{ij}\,\dot \gamma_{ij}+\Omega \,\dot 
{\bar t}+\Omega_i\, {\dot {\bar \xi}}^i \\
\nonumber\!\!&-&\!\!\mathcal{H}''_c-u\,\Omega-u^i\Omega_i-v\,\tilde 
\chi-v^i\tilde \chi_i-w \,\bar{\mathcal{H}}-w^i\bar{\mathcal{H}_i}\Big]\,,  
\end{eqnarray}
where 
\begin{eqnarray}\label{5Hamiltoni56}
\mathcal{H}''_c&=&\alpha
\,\bar{\mathcal{H}}+\alpha^i\,\mathcal{H}_i-\frac{1}{2}\,\left(\alpha^2\gamma\right)_{,i}\gamma^{ij}\,\tilde
\chi_j-\frac{d-3}{4(d-2)}\,\alpha^2\gamma\,\gamma^{ij}\,\tilde
\chi_i\tilde \chi_j-\alpha^j\alpha^i_{,j}\tilde
\chi_i\\ \nonumber &+& \frac{1}{\alpha}\,\alpha^i\alpha^j\alpha_{,j}\tilde
\chi_i-2\,\alpha\alpha^k\gamma_{jk}\pi^{ij}\tilde
\chi_i+\frac{1}{d-1}\,\alpha\alpha^i\gamma_{jk}\pi^{jk}\tilde\chi_i+\frac{1}{d-1}\,\frac{\alpha^i\alpha_{,i}}{\alpha}\tilde\chi\\
\nonumber &+& \alpha^2\alpha^i\pi\tilde\chi_i+\alpha\alpha^i\alpha^j\pi_i\tilde\chi_j+\alpha\alpha^l\pi_l\tilde\chi-\frac{2}{d-1}\,\alpha\gamma_{jk}\pi^{jk}\tilde \chi+\frac{d-2}{d-1}\left(\alpha^2\pi+\alpha^l_{,l}\right)\tilde\chi\,.
\end{eqnarray}
By applying the reduction method of Faddeev and Jackiw
\cite{5Faddeev1988} to the action of eq. (\ref{5action32}), one obtains
\begin{eqnarray}\label{5action321}
S=\int dx \,\Big[\,\pi^{ij}\,\dot \gamma_{ij}-\alpha \,\bar{\mathcal{H}}-\alpha^i\,\mathcal{H}_i-w \,\bar{\mathcal{H}}-w^i\bar{\mathcal{H}_i}\Big]\,,  
\end{eqnarray}
upon dropping surface terms. This variant of
the ADM action has been used by Teitelboim
\cite{5Teitelboim1982,5Teitelboim1983} and Ashtekar \cite{5Ashtekar1987} in
quantum gravity, and by York \emph{et.al.} in numerical relativity
\cite{5Abrahams1998,5Choquet1998}. Since the 
constraints $\mathcal{H}_i$ and $\bar{\mathcal{H}}$ are  
derived from the constraints $\tilde \tau_i$ and $\tilde \tau$ of
eqs. (\ref{5ax51}) and (\ref{5ax55}) under the canonical transformations of
eqs. (\ref{5gamma}) and (\ref{5cantran22}) as in
eqs. (\ref{5axtrauma}) and (\ref{5axtrauma2}),
the algebra of the PB of these constraints is
\begin{eqnarray}\label{5algebra5}
f\big\{\mathcal{H}_i,\mathcal{H}_j\big\}g&=&gf_{,j}\mathcal{H}_i-fg_{,i}\mathcal{H}_j\,,\\ \label{5algebra52}
f\big\{\bar{\mathcal{H}},\bar{\mathcal{H}}\big\}g&=&\left(gf_{,i}-fg_{,i}\right)\gamma\,\gamma^{ij}\,\mathcal{H}_j\,,\\
\label{5algebra53}
f\big\{\mathcal{H}_i,\bar{\mathcal{H}}\big\}g &=&
\left(gf_{,i}-fg_{,i}\right)\bar{\mathcal{H}}\,,
\end{eqnarray}
according to eqs. (\ref{5ax60}), (\ref{5ax66}) and
(\ref{5ax67}), consistent with the constraint algebra given in
\cite{5Teitelboim1982,5Teitelboim1983}. (Here $f$ and $g$ are test functions.)  

\section{Tentative gauge constraints}\label{iii}
Together with a set of ``admissible'' gauge constraints, one
may put the first class constraints of the extended action (which are now second class)
strongly equal to zero and solve them in order to eliminate the
redundant degrees of freedom from the action and introduce the
DB. Meanwhile, \emph{all} the gauge freedom of the Lagrangian
action is fixed.

We now consider gauge fixing conditions for the action of
eq. (\ref{5action}). A study of the equations of motion of the
extended action of eq. (\ref{5action}) is
illuminating in the nature and role of the canonical variables
employed in this action. If we are only
interested in the equations of motion derived from this action we may
then rewrite it as 
\begin{eqnarray}\label{5action-re}
S=\int dx \,\Big[\,\Omega\, \dot{\bar t}+\Omega_i\,\dot{{\bar \xi}}^i+\,\Pi
\,\dot h &\!\!+\!\!&\Pi_i\, \dot
h^i+\Pi_{ij}\,\dot q^{ij}\\ \nonumber &\!\!-\!\!&\bar u\,\Omega-{\bar
u}^i\,\Omega_i-\bar v\,\tilde 
\chi-{\bar v}^i\tilde \chi_i-\bar w \,\tilde \tau-{\bar w}^i\tilde \tau_i\Big]\,, 
\end{eqnarray} 
where we have shifted the Lagrange multipliers by adding to them
the coefficients of the constraints appearing in the Hamiltonian $\mathcal{H}_c$ of
eq. (\ref{5Hamiltoni}). The equations of motion for $\bar u$, ${\bar
u}^i$, $\bar t$ and
$\bar \xi^i$ are trivially satisfied while the equations of motion for
$\Omega$ and $\Omega_i$ show that $\bar t$ and ${\bar \xi}^i$ are undetermined, 
\begin{eqnarray}\label{tetso}
\dot{\bar t}&\approx&\bar u\,,\\ \nonumber
\dot {\bar\xi}^i&\approx&\bar u^i\,.
\end{eqnarray}    
Therefore, tentative gauge constraints for the primary first class
constraints  
\begin{eqnarray}\label{gaucon0}
\Omega&\approx&0\,,\\ \nonumber 
\Omega_i&\approx&0\,,
\end{eqnarray} 
could be of the form
\begin{eqnarray}\label{gaucon01}
\bar t-C_{\bar t}(x)&\approx& 0\,,\\ \nonumber 
\bar\xi^i-C_{\bar \xi^i}(x)&\approx&0\,,
\end{eqnarray}
respectively, where $C_{\bar t}$ and $C_{\bar \xi^i}$ are arbitrary
functions.  The constraints of
eqs. (\ref{gaucon0},\ref{gaucon01}) form a minimal set of second class
constraints and may thus be turned into strong equations. The DB of
the rest of the canonical variables remains their PB. 
We now prove that much like $\bar t$ and $\bar \xi^i$, the fields $h$ and
$h^i$ are left undetermined by the equations of motion. By extremizing
the action of eq. (\ref{5action-re}), the equations of motion corresponding to $\Pi$,
$\Pi_i$, $h$, $h^i$, $\bar v$ and ${\bar v}^i$ are
\begin{eqnarray}\label{forh}
\frac{\delta S}{\delta \Pi}&=&\dot h-\bar v h=0\,,\\ \label{forh2}
\frac{\delta S}{\delta \Pi_i}&=&\dot {h^i}+{\bar v}^i h=0\,,\\ \label{forh3}
\frac{\delta S}{\delta h}&=&-\,\dot \Pi-\bar v\,\Pi+{\bar v}^i\Pi_i+{\bar v}^i_{,i}=0\,,\\ \label{forh4}
\frac{\delta S}{\delta h^i}&=&-\,\dot{\Pi_i}+{\bar v}_{,\,i}=0\,,\\
\label{forh5}
\frac{\delta S}{\delta {\bar v}}&=&h^l_{,\,l}+h\,\Pi=0\,,\\ \label{forh6}
\frac{\delta S}{\delta {\bar v}^i}&=&h_{,\,i}-h\,\Pi_i=0\,.
\end{eqnarray}
In obtaining eqs. (\ref{forh}-\ref{forh4}) we
have used the constraint equations (\ref{forh5},\ref{forh6}). Since
the Lagrange multipliers $\bar v$ and  
${\bar v}^i$ are arbitrary, the fields $h$ and $h^i$ can 
take the values of any arbitrary functions $C_h(x)$ and
$C_{h^i}(x)$, as justified below. Suppose the latter is true, that is, $h=C_h(x)$ and
$h^i=C_{h^i}(x)$. Eqs. (\ref{forh}), (\ref{forh2}), (\ref{forh5}) and
(\ref{forh6}) may be solved for $\bar v$, $\bar v^i$, $\Pi$ and
$\Pi_i$ in order to express them in terms of $C_h(x)$ and
$C_{h^i}(x)$. Upon substituting these solutions into eqs. (\ref{forh3})
and (\ref{forh4}) they result in trivial identities. 

The foregoing observation suggests
that tentative gauge constraints corresponding to the secondary first
class constraints 
\begin{eqnarray}\label{gaucon1}
\tilde \chi&\equiv& h^l_{,\,l}+h\,\Pi\approx0\,,\\ \nonumber
\tilde\chi_i&\equiv& h_{,\,i}-h\,\Pi_i\approx0\,,
\end{eqnarray}  
and {\it{compatible with the equations of motion}} could be of the form
\begin{eqnarray}\label{gaucon11}
h-C_h(x)&\approx&0\,,\\ \nonumber 
h^i-C_{h^i}(x)&\approx&0\,,
\end{eqnarray} 
where $C_h$ and $C_{h^i}$ are arbitrary functions. Once again, the constraints of
eqs. (\ref{gaucon1},\ref{gaucon11}) form a minimal set of second class
constraints which may be turned into strong equations. Once the
solutions of these equations are inserted into the action of eq. (\ref{5action}) it is
reduced to   
\begin{eqnarray}\label{5action-rere}
S=\int dx \,\Big[\,\Pi_{ij}\,\dot q^{ij}- {\bar{w}} \,\tilde
\tau-{{\bar w}}^i\tilde \tau_i\Big] \,,
\end{eqnarray} 
upon dropping an irrelevant surface term. (The redefined Lagrange
multipliers $\bar w$ and ${\bar w}^i$ are arbitrary and can depend on $q^{ij}$ and
$\Pi_{ij}$). Since the constraints of
eqs. (\ref{gaucon1},\ref{gaucon11})  do not involve $q^{ij}$ and $\Pi_{ij}$, the PB
of these variables remains unchanged upon solving the constraints of
eqs. (\ref{gaucon1},\ref{gaucon11}) and introducing the DB. We note
that the functions $C_{\bar t}$, $C_{{\bar \xi}^i}$, $C_h$ and
$C_{h^i}$ can depend on $q^{ij}$ and $\Pi_{ij}$ without
violating any of the arguments and conclusions made above, since under
such an assumption the constraints of eqs. (\ref{gaucon1},\ref{gaucon11})
are proven to be of special form as follows. If 
$$\left\{\theta_s\right\}=\left\{h-C_h(\gamma_{ij},\pi^{ij}),h^k-C_{h^k}(\gamma_{ij},\pi^{ij}),\tilde\chi,\tilde\chi_k\right\}\,,$$
we have
\begin{eqnarray}\label{55matrix}
\left\{\theta,\theta\right\}^{-1}\approx(1/C_h)^2
\begin{pmatrix}
0&0&C_h&0\\
0&0&0&-\frac{1}{d-1}\,\delta^i_j\,C_h\\
-C_h&0&\left\{C_h,C_h\right\}&\left\{C_{h^i},C_h\right\}\\
0&\frac{1}{d-1}\,\delta^i_j\,C_h&\left\{C_h,C_{h^j}\right\}&\left\{C_{h^k},C_{h^l}\right\}
\end{pmatrix}\,,
\end{eqnarray} 
which implies that the DB of $q^{ij}$ and $\Pi_{ij}$ remains
equal to their PB upon turning
the first class constraints $\tilde \chi$ and $\tilde \chi_i$ and
their corresponding gauge constraints into strong equations, thanks to
the constraints $\tilde \chi$ and $\tilde \chi_i$ not depending on
$q^{ij}$ and $\Pi_{ij}$. The action of eq. (\ref{5action-rere}),
therefore, is identical with the Faddeev  
action of eq. (\ref{5Faddeev}) upon appropriate gauge fixation.

The gauge constraints of eqs. (\ref{gaucon01}) and (\ref{gaucon11})
are not in general admissible for arbitrary functions $C_{\bar t}$,
$C_{{\bar \xi}^i}$, $C_h$ and $C_{h^i}$, since they can not be achieved from an
arbitrary configuration of the fields $\bar t$, $\bar \xi^i$, $h$ and $h^i$ by a
diffeomorphism invariance transformation. In principle, one needs to
consider the gauge constraints corresponding to the tertiary
constraints $\bar \tau$ and $\bar \tau_i$ along with the gauge
constraints of eqs. (\ref{gaucon01}) and (\ref{gaucon11}), and choose
appropriate functions $C_{\bar t}$, $C_{\bar \xi^i}$, $C_h$ and
$C_{h^i}$ in such a way that the gauge constraints altogether are achieved by diffeomorphism
invariance transformations, while in this process the gauge functions
are completely fixed upon assuming appropriate behavior of the gauge
functions on the boundaries. 

More insight about eqs. (\ref{forh}-\ref{forh6}) and the role of the
fields $h$, $h^i$, $\Pi$ and $\Pi_i$ in the action of
eq. (\ref{5action}) can be gained in the following way. We may add a
surface term of the form
\begin{eqnarray}\label{suurf}
\mathcal{S}=-\left(\frac{h^lh_{,l}}{h}\right)_{,0}+\left(\frac{\dot h h^l}{h}\right)_{,l}
\end{eqnarray} 
to the kinetic part of the action of eq. (\ref{5action}) and write
it as
\begin{eqnarray}\label{suurf2}
\Pi\,\dot h+\Pi_i\,\dot h^i+\Pi_{ij}\dot
q^{ij}+\mathcal{S}&=&\bar \Pi \,\dot h+\bar \Pi_i \,\dot
h^i+\Pi_{ij}\dot q^{ij}
\end{eqnarray}
where
\begin{eqnarray}\label{barchi-1}
\bar \Pi&=&\frac{1}{h}\,\tilde \chi,\\\label{barchi-2}
\bar \Pi_i&=&-\,\frac{1}{h}\,\tilde\chi_i\,,
\end{eqnarray}
with $\tilde \chi$ and $\tilde \chi_i$ given by
eqs. (\ref{5ax22},\ref{5ax23}). In particular
\begin{eqnarray}\label{barcom}
\left\{\bar\Pi,\bar\Pi\right\}=\left\{\bar
\Pi,\bar\Pi_i\right\}=\left\{\bar \Pi_i,\bar \Pi_j\right\}=0
\end{eqnarray}
according to eqs. (\ref{5ax24}-\ref{5ax26}). We may therefore observe
that $\bar \Pi$ and $\bar \Pi_i$ are the momenta conjugate to $h$
and $h^i$, and write the action of eq. (\ref{5action}) as
\begin{eqnarray}\label{5actionbar}
S=\int dx \,\Big[\,\bar \Pi \,\dot h\!\!&+&\!\!\bar \Pi_i\, \dot
h^i+\Pi_{ij}\,\dot q^{ij}+\Omega \,\dot {\bar 
t}+\Omega_i\, {\dot {\bar \xi}}^i \\
\nonumber\!\!\!\!&-&\!\!\mathcal{H}_c-u\,\Omega-u^i\,\Omega_i-v\,\bar 
\Pi-v^i\bar \Pi_i-w \,\tilde \tau-w^i\tilde \tau_i\Big] 
\end{eqnarray} 
where now
\begin{eqnarray}\label{5Hamiltonibar}
\mathcal{H}_c&=&\frac{1}{h}
\,\tilde \tau+\frac{h^i}{h}\,\tilde
\tau_i+\frac{1}{2(d-2)}\,\frac{1}{h}\,q_{kl}q^{kl}_{\,\,,i}\,q^{ij}\bar
\Pi_j-\frac{d-3}{4(d-2)}\,q^{ij}\bar \Pi_i\bar
\Pi_j\\ \nonumber &-&\frac{h_{,j}}{h^2}\,q^{ij}\bar \Pi_i+\frac{h^j}{h}\,h^i_{,j}\bar
\Pi_i-2\,\frac{h^m}{h^2}\,q^{il}\Pi_{lm}\bar
\Pi_i+\frac{h^i}{h}\,h_{,i}\,\bar\Pi+\left(\bar
\Pi-\frac{1}{h}\,h^l_{,l}\right)h^i\bar\Pi_i\\ \nonumber
&+&\frac{h^i}{h}\,q^{mn}\Pi_{mn}\,\bar
\Pi_i-\frac{h^l}{d-1}\,\left(\frac{1}{h}\,h^l_{,l}+\bar \Pi_l\right)\bar
\Pi+\frac{d-2}{d-1}\left(h\bar \Pi +2\,q^{ij}\Pi_{ij}
-2\,h^l_{,l}\right)\bar \Pi \\ \nonumber &-& 
\frac{\bar t}{d-1}\,h \,\bar \Pi+\,\bar \xi^i(h\bar \Pi_i)+\tilde B^i\,\tilde \Lambda_i+\tilde B^{ij}\, \tilde \Lambda_{i} \tilde \Lambda_j\,,
\end{eqnarray}
Written in this from, it is explicitly seen that the fields $h$ and
$h^i$ act as Lagrange multiplier fields, much in the same way as the fields
$\bar t$ and $\bar \xi^i$ are Lagrange multipliers. Such a simplification of the action is
reminiscent of Dirac's simplification of the Hamiltonian formulation of
the second order EH action by addition of the following surface terms to the EH
Lagrangian \cite{5Dirac1958}
\begin{eqnarray}\label{Diracsurface}
\left[\left(\sqrt{-\mathfrak{g}}\,g^{00}\right)_{,\nu}\frac{g^{\nu
0}}{g^{00}}\right]_{,0}-\left[\left(\sqrt{-\mathfrak{g}}\,g^{00}\right)_{,0}\frac{g^{\nu
0}}{g^{00}}\right]_{,\nu}\,, 
\end{eqnarray}
resulting in the primary constraints taking the simple
form
\begin{eqnarray}\label{Diracprimar}
p^{\mu0}=0\,;
\end{eqnarray}
in contrast with the second order Hamiltonian
formulation of Pirani and Schild \cite{5Pirani1952} in which
the EH action is considered without these surface terms, and
the primary constraints are of the more complicated form $p^{\mu
0}=p^{\mu0}(\bar q,\bar p)$, with $\bar q$
and $\bar p$ being other canonical variables. (The two approaches have
been compared and contrasted in 
\cite{5Frolov}.) The surface terms of eq. (\ref{Diracsurface}) indeed reduce to the
surface terms of eq. (\ref{suurf}).    

Since the gauge constraints of eqs. (\ref{gaucon01}) and
(\ref{gaucon11}) are canonical, one may use them in order to fix the
gauge freedom of the actions of eqs. (\ref{5action3}) and
(\ref{5action32}) if they are transformed 
under the associated canonical transformations. In the case of the action of
eq. (\ref{5action}) when written in terms of $N$, $N^i$,
$\gamma_{ij}$ and their conjugate momenta $p$, $p_i$ and $p^{ij}$
defined in eqs. (\ref{5tootra}-\ref{5tootra3}), a reduction to the ADM
action is not quite immediate. In particular, since the
constraints $\tilde \chi$ and 
$\tilde \chi_i$ of eqs. (\ref{5totaltra},\ref{5totaltra2}) depend on
$\gamma_{ij}$, we expect the PB of $\gamma_{ij}$ and $\pi^{ij}$ to be altered
upon solving the constraints $\tilde \chi$ and $\tilde \chi_i$ and
introducing the DB if the gauge
constraints $C_h$ and $C_{h^i}$ depend on $\gamma_{ij}$ and
$p^{ij}$. For the specific class of admissible
gauge constraints in which $N$ and $N^i$ are constant ($N=1$ and
$N^i=0$ for instance) a reduction to a ``gauge-fixed'' ADM
action is seen to easily be realized. A more straightforward reduction to the ADM action might be
possible if we assume that the gauge constraints also depend on the
momenta $p$ and $p_i$.
  
\section{Gauge transformations}\label{ii}
When written in terms of $q^{ij}$ or $\gamma_{ij}$, the problem of
determining the gauge transformations of the first order EH Lagrangian
action from the first class constraints generated in the Hamiltonian
formulation transforms into a more manageable task than when one works with
the formalism in which $H^{ij}$ is used. This simplification
occurs mainly because in terms of the former variables constraints of
different stage depend on different sets of the canonical variables, as explained in
previous sections. In
this section we consider the action of eq. (\ref{5action}) (which is a
functional of $h$, $h^i$, $q^{ij}$ and their conjugate momenta) and derive
the explicit form of the generator of the gauge transformations of the
fields $h$, $h^i$, $q^{ij}$, $\Pi$, $\Pi_i$ and $\Pi_{ij}$. The gauge
transformations of $\bar t$ and $\bar \xi^i$ which 
act as Lagrange multipliers are given by separate equations which are
necessary for the action to remain invariant under the gauge
transformations.  This is done using a method very similar to the
method of HTZ \cite{5Henneaux1992}. In this approach one directly considers gauge
transformations of the total action instead of the gauge
transformations that leave the extended action invariant \cite{5Sundermeyer}.
Using the generator thus obtained, we explicitly evaluate the gauge transformation of the field
$h=\sqrt{-\mathfrak{g}}\,g^{00}$ assuming the gauge functions
corresponding to the tertiary constraints to be independent of the
canonical variables, and show that a field dependent
redefinition of the gauge functions is necessary in order for this
transformation to correspond to the usual diffeomorphism
invariance, which is given by \cite{5Faddeev1975}  
\begin{eqnarray}\label{gaugetrt}
\delta\,h^{\mu\nu}=-\left(h^{\mu\nu}\eta^\lambda\right)_{\!,\,\lambda}+h^{\mu\lambda}\,\eta^\nu_{,\lambda}+h^{\nu\lambda}\,\eta^\mu_{,\lambda}\,, 
\end{eqnarray}
for the fields $h^{\mu\nu}=\sqrt{-\mathfrak{g}}\,g^{\mu\nu}$, where
$\eta^\mu$ are arbitrary descriptors \cite{5Bergmann}.

It has been shown that for most relevant field theories one may drop fields (and
their corresponding momenta) that act as Lagrange multipliers from the
total Hamiltonian without loss of the gauge transformations if after
the emilination the Lagrange multipliers are identified
with the eliminated coordinates \cite{5Mukhanov}.  We therefore rewrite the
total action corresponding to the extended action of eq. (\ref{5action}) as 
\begin{eqnarray}\label{5actiont}
S_T=\int dx \,\Big[\,\Pi \,\dot h+\Pi_i\, \dot
h^i+\Pi_{ij}\,\dot q^{ij}-\mathcal{H}_T\Big]\,, 
\end{eqnarray}  
where 
\begin{eqnarray}\label{Tota}
\mathcal{H}_T&=&\frac{1}{h}\, \tau+\frac{h^i}{h}\,
\tau_i-t\,\chi-\xi^i\chi_i\,,\\ \label{Tota2}
t&=&\frac{\bar
t}{d-1}-\frac{h^i}{h^2}\,h_{,i}+\frac{1}{d-1}\,\frac{h^l}{h}\,\Pi_l-\frac{d-2}{d-1}\,\frac{1}{h}\,\left(h\Pi+2\,q^{ij}\Pi_{ij}-h^l_{,l}\right)\\\label{Tota3}
\xi^i&=&\bar
\xi^i+\frac{1}{2(d-2)}\,\frac{1}{h^2}\,q_{kl}q^{kl}_{\,\,,j}\,q^{ij}+\frac{d-3}{4(d-2)}\,\frac{1}{h^2}\,q^{ij}\tilde
\chi_j-\frac{h_{,j}}{h^3}\,q^{ij}\\ \nonumber &+&\frac{h^j}{h^2}\,h^i_{,j}
-2\,\frac{h^m}{h^2}\,q^{il}\Pi_{lm}+\frac{h^i}{h}\,\Pi+\frac{h^i}{h^2}\,q^{mn}\Pi_{mn}\,. 
\end{eqnarray}
(Note that we have dropped the tilde from the constraints of
eqs. (\ref{5ax22},\ref{5ax23},\ref{5ax51},\ref{5ax55-2}).) The
usefulness of the redefinitions of eqs. (\ref{Tota2},\ref{Tota3}) 
lies in that the terms other than $\bar t$ and $\bar \xi^i$ which are
included in $t$ and $\xi^i$ do not contribute 
to the gauge transformations of the fields $h$, $h^i$, $q^{ij}$ and
their conjugate momenta $\Pi$, $\Pi_i$ and $\Pi_{ij}$  but to the gauge
transformations of $\bar t$ and $\bar \xi^i$ which now explicitly appear as the
Lagrange multiplier fields. We emphasize that 
the actual dependence of $t$ and $\xi^i$ of
eqs. (\ref{Tota2},\ref{Tota3}) on the canonical variables is 
quite important for obtaining the gauge transformations of $\bar t$
and $\bar \xi^i$.

In contrast with the first and second order formulations of the free spin
two field actions considered in \cite{5Ghalati2007-2,5Ghalati2008}, in
which the structure functions were constant, we need 
to consider a more general formalism when dealing with the gauge
transformations of the full EH action, where we need to consider the structure functions
to be field dependent. The most general form of the generator $G$ of the gauge
transformations of the total action of eq. (\ref{5actiont}) is
\begin{eqnarray}\label{gen5}
G=\int dx \left(\bar \mu \,\chi+{\bar \mu}^i\chi_i+\mu\,\tau+\mu^i\tau_i\right)\,,
\end{eqnarray}  
where the gauge functions $\mu$ and
$\mu^i$ corresponding to the tertiary constraints $\tau$ and
${\tau}_i$ are arbitrary
functions depending on spacetime as well as the canonical 
variables, and the functions $\bar \mu$ and ${\bar \mu}^i$ are
arbitrary functions of spacetime and the canonical variables which satisfy a set
of differential equations that arise by requiring the invariance of the
total action.\footnote{We consider the special case where the gauge
functions do not depend on the Lagrange multiplier fields.} 
Using eq. (\ref{gen5}) we may show that
\begin{eqnarray}\label{delta5}
\delta H_T=-\left({\bar \delta \chi}+{\bar \delta}^i\chi_i+\delta \tau+\delta^i\tau_i\right)
\end{eqnarray}
where
\begin{eqnarray}\label{deltas5}
\bar \delta
\chi=\int\!\!dx' dx\, \Bigg[\!\!\!&\chi&\!\!\!\bigg(\bigg\{t,\chi\bigg\}\,\bar
\mu+\bigg\{t,\chi_i\bigg\}\,{\bar
\mu}^i+\bigg\{t,\tau\bigg\}\,\mu+\bigg\{t,\tau_i\bigg\}\,\mu^i\bigg)\\
\nonumber&-&\bigg\{\mathcal{H}_T,\bar\mu\bigg\}\,\,\,\chi\,\,\Bigg]
\end{eqnarray}
\begin{eqnarray}\label{deltas5-2}
{\bar \delta}^i\chi_i=\int\!\! dx' dx \Bigg[\!\!\!&\chi_i&\!\!\!\,\bigg(\bigg\{\xi^i,\chi\bigg\}\,\bar
\mu+\bigg\{\xi^i,\chi_j\bigg\}\,{\bar
\mu}^j+\bigg\{\xi^i,\tau\bigg\}\,\mu+\bigg\{\xi^i,\tau_j\bigg\}\,\mu^j\bigg)\\
\nonumber &-&\bigg\{\mathcal{H}_T,{\bar \mu}^i\bigg\}\,\,\chi_i\,\,\Bigg]
\end{eqnarray}
\begin{eqnarray}\label{deltas5-3}
\delta\tau=\int dx \Bigg[\Bigg(\frac{1}{h}\,\bar \mu+\frac{1}{h}\,\mu^i_{,i}-\left(\frac{h^i}{h}\right)_{\!\!,i}\!\!\mu^i-\left(\frac{h^i}{h}\right)_{\!\!,i}\!\!\mu+\frac{h^i}{h}\,\mu_{,i}-\bigg\{H_T,\mu\bigg\}\Bigg)\,\,\tau\,\,\Bigg]
\end{eqnarray}
\begin{eqnarray}\label{deltas5-4}
\delta^i\tau_i=\int dx
\Bigg[\Bigg({\bar \mu}^i\!\!\!&-&\!\!\!\,q^{ij}\bigg(\frac{1}{h}\bigg)_{\!\!,j}\!\!\mu+\frac{1}{h}\,q^{ij}\mu_{,j}+\frac{h^i}{h}\,\bar 
\mu-\left(\frac{h^i}{h}\right)_{\!\!,j}\mu^j+\frac{h^j}{h}\,\mu^i_{,j}\\ \nonumber &-& \bigg\{H_T,\mu^i\bigg\}\Bigg)\,\,\tau_i\,\,\Bigg]\,.
\end{eqnarray}
Since 
\begin{eqnarray}\label{kinetic}
\delta \int dx\left(\Pi\,\dot h+\Pi_i\,\dot h^i+\Pi_{ij}\,\dot
{q^{ij}}\right)&=&\int dx\left({\bar \mu}_t\,\chi+{{\bar
{\mu}}_t}^i\,\chi_i+ \mu_t\, \tau+ \mu_t^i\,\tau_i\right)\,,
\end{eqnarray}
where the partial derivative with respect to time is denoted by a $t$
index, we then have 
\begin{eqnarray}\label{actionvar}
\delta S_T=\int dx \Bigg[\left({\bar
\mu}_t+\bar \delta.+\frac{\delta
t}{d-1}\,\right)\,\chi\!\!\!&+&\!\!\!\bigg({\bar\mu}_t^i+\bar
\delta^i\!.+\delta \bar \xi^i\bigg)\,\chi_i \\ \nonumber
&+&\!\!\bigg(\mu_t+\delta.\bigg)\tau+\bigg(\mu_t^i+\delta^i\!.\bigg) 
\tau_i\Bigg]\,,
\end{eqnarray}
where we have symbolically written $\bar \delta\chi=\bar \delta\,.\,
\chi$\,, etc. to indicate that the integral signs have been dropped
after all PBs have been evaluated and the derivatives over the constraints
$\chi$, $\chi_i$, $\tau$ and $\tau_i$ have been removed by addition of
appropriate surface terms. If we require
the total action of eq. (\ref{5actiont}) to be invariant under 
the gauge transformations of eq. (\ref{gen5}) we have $\delta
S_T=0$, which is satisfied only if the coefficients of the constraints $\chi$,
$\chi_i$, $\tau$ and $\tau_i$ are set equal to zero. By a choice of the gauge
functions $\mu$ and $\mu^i$ corresponding to the tertiary constraints
$\tau$ and $\tau_i$, we may determine the gauge functions $\bar \mu$
and $\bar \mu^i$ corresponding to the secondary constraints $\chi$ and
$\chi_i$ by setting the coefficients of $\tau$ and $\tau_i$ in
eq. (\ref{actionvar}) equal to zero. In particular, we note that
according to eqs. (\ref{deltas5-3},\ref{deltas5-4}) these are simple
algebraic equations for the gauge functions $\bar \mu$ and $\bar
\mu^i$. Vanishing of the coefficients of the constraints $\chi$ and
$\chi_i$ in eq. (\ref{actionvar}), on the other hand, provides with
the gauge transformations of the Lagrange multipliers $\bar t$ and
$\bar \xi^i$.

Let us choose the gauge functions $\mu$ and $\mu^i$ to depend only on
spacetime and not on the canonical variables,
\begin{eqnarray}\label{gfun}
\mu=\mu(x)\qquad\qquad\qquad\mu^i=\mu^i(x)\,.
\end{eqnarray}
This choice is not necessary in principle, and one may choose any
arbitrary functions that depend on the canonical variables as
well. Setting the coefficients of the constraints $\tau$ and $\tau_i$
in eq. (\ref{actionvar}) equal to zero and solving for $\bar \mu$ and $\bar \mu^i$
using $\mu$ and $\mu^i$ of eq. (\ref{gfun}) gives,
\begin{eqnarray}\label{change5}
\bar \mu&=&-h\bigg(\dot
\mu+\frac{1}{h}\,\mu^i_{,i}-\left(\frac{1}{h}\right)_{\!\!,i}\mu^i-\bigg(\frac{h^i}{h}\bigg)_{\!\!,i}\mu+\frac{h^i}{h}\,\mu_{,i}\bigg)\\
\label{change5-2}
{\bar\mu}^i&=&-\dot
\mu^i+h^i\dot\mu +q^{ij}\bigg(\frac{1}{h}\bigg)_{\!\!,j}\mu-\frac{1}{h}\,q^{ij}\mu_{,j}+\bigg(\frac{h^i}{h}\bigg)_{\!\!,j}\mu^j-\frac{h^j}{h}\,\mu^i_{,j}+\frac{h^i}{h}\mu^j_{,j}\\
\nonumber &-&h^i\bigg(\frac{1}{h}\bigg)_{\!\!,j}\,\mu^j-h^i\bigg(\frac{h^j}{h}\bigg)_{\!\!,j}\mu+\frac{h^ih^j}{h}\,\mu_{,j}\,,
\end{eqnarray}
where in obtaining eq. (\ref{change5-2}) we have used
eq. (\ref{change5}). The generator of gauge transformations $G$ is
therefore given by eq. (\ref{gen5}), with $\mu$, $\mu^i$, $\bar \mu$ and $\bar
\mu^i$ given by eqs. (\ref{gfun}-\ref{change5-2}). Using this
generator we may find the gauge transformations of $h$, $h^i$,
$q^{ij}$, $\Pi$, $\Pi_i$ and $\Pi_{ij}$. The gauge
transformation for $h$ is thus
\begin{eqnarray}\label{gaugeh}
\delta h&=&\left\{h,G\right\}\\ \nonumber 
&=&-h^2\dot\mu-h\,\mu^i_{,i}-h_{,i}\,\mu^i-h^ih_{,i}\,\mu+h\,h^i_{,i}\,\mu-h\,h^i\mu_{,i}\,.
\end{eqnarray}
This is identical with the diffeomorphism invariance transformation of
$h$ given by eq. (\ref{gaugetrt}) if we substitute 
\begin{eqnarray}\label{trt}
\eta^0&=&-h\,\mu\,,\\ \label{trt2}
\eta^i&=&\mu^i-h^i\mu\,, 
\end{eqnarray}
for the descriptor $\eta^\mu$ in eq. (\ref{gaugetrt}). 

The gauge transformations of the fields $h^i$, $q^{ij}$, $\Pi$, $\Pi_i$ and
$\Pi_{ij}$ can be determined using the gauge generator $G$ of
eq. (\ref{gen5}). One may thus easily observe that by the
dependence of the constraints $\chi$ and $\chi_i$ on the derivatives of
$h$ and $h^i$ the gauge transformations for $\Pi$ and $\Pi_i$ involve
first order derivatives of $\bar \mu$ and $\bar \mu^i$ and thus
second order derivatives of $\mu$ and $\mu^i$. Also, since
$\tau$ depends on second order derivatives of the fields $q^{ij}$,
we see how second order derivatives of the gauge functions $\mu$
and $\mu^i$ enter the gauge transformations of $\Pi_{ij}$. The
gauge transformations for the Lagrange  
multiplier fields $\bar t$ and $\bar \xi^i$ on the other hand are
obtained by requiring that the coefficients of 
the constraints $\chi$ and $\chi_i$ in eq. (\ref{actionvar}) vanish,
which according to eqs. (\ref{actionvar},\ref{change5},\ref{change5-2}) involve
second order derivatives of the gauge functions $\mu$ and $\mu^i$. The
existence of second order derivatives of the gauge functions $\mu$ and
$\mu^i$ is expected for the gauge invariance of the fields $\Pi$,
$\Pi$, $\Pi_i$, $\Pi_{ij}$, $\bar t$ and $\bar \xi^i$ produced by the
gauge generator $G$ of eq. (\ref{gen5}) to coincide with their
diffeomorphism invariance, which is found by the diffeomorphism
invariance of the Christoffel symbols \cite{5Faddeev1975,5Faddeev1982}.  

We have verified that if we had used the action of eq. (\ref{5action})
instead of the action of eq. (\ref{5actiont}) for evaluation of the gauge
trasnformations, we would have obtained gauge symmetries which
differed from the gauge symmetries obtained above by trivial equations of motion
symmetries. Such symmetries have been discussed in \cite{5Henneaux1992}.   

\section{Linearized Theory}\label{XV}

The Linearized theory of the novel Hamiltonian formulation of the
extended EH action of eq. (\ref{action}) can
be obtained by linearizing the fields $h$, $h^i$ and 
$H^{ij}$ around the metric of the flat spacetime,
\begin{eqnarray}\label{5lin}
h^{\mu\nu}&=&\eta^{\mu\nu}+\tilde h^{\mu\nu}\,,
\end{eqnarray}
where the signature of the metric of the flat spacetime is
$\eta^{\mu\nu}=(-,+,+,\ldots,+)$, and  we have ignored terms of higher
order in ${\tilde h}^{\mu\nu}$. This implies that, in particular,
\begin{eqnarray}\label{5lin2}
h=-1+\tilde h,\quad\quad h^i=\tilde h^i,\quad\quad H^{ij}=-\delta^{ij}-\tilde
h^{ij},\quad\quad H_{ij}=-\delta_{ij}+\tilde h_{ij}
\end{eqnarray} 
if we keep terms linear in the perturbation fields only. Under the
expansion of eq. (\ref{5lin2}), the fundamental PBs
\begin{eqnarray}\label{5pbfun}
\left\{h,\omega\right\}=1\,,\qquad\,\,\,\,\,\,\left\{h^i,\omega_j\right\}=\delta^i_j\,,\qquad\,\,\,\,\,\,\left\{H^{ij},\omega_{kl}\right\}=\frac{1}{2}\,\left(\delta^i_k\delta^j_l+\delta^j_k\delta^i_l\right), 
\end{eqnarray} 
transform into
\begin{eqnarray}\label{5pbfun2}
\left\{\tilde
h,\omega\right\}=1\,,\qquad\,\,\,\,\left\{{\tilde{h}}^i,\omega_j\right\}=\delta^i_j\,,\qquad\,\,\,\,\left\{{\tilde 
h}^{ij},-\,\omega_{kl}\right\}=\frac{1}{2}\,\left(\delta^i_k\delta^j_l+\delta^j_k\delta^i_l\right), 
\end{eqnarray}
showing that the fields $\omega$, $\omega_i$ and $-\,\omega_{ij}$
act as the momenta conjugate to the perturbation fields $\tilde h$,
${\tilde h}^i$ and ${\tilde h}^{ij}$. Keeping only terms in the EH Hamiltonian action of
eq. (\ref{action}) which are bilinear in the fields and Lagrange
multipliers, and by defining  
\begin{eqnarray}\label{5def}
\tilde \omega_{ij}=-\,\omega_{ij}
\end{eqnarray}
we obtain
\begin{eqnarray}\label{action2}
S=\int dx \,\Big[\,\omega \,\dot {\tilde{h}}\!\!&+&\!\!\omega_i\, \dot
{\tilde h}^i+\tilde {\omega}_{ij}\,\dot {\tilde{h}}^{ij}+\Omega \,\dot{ 
\bar t}+\Omega_i\, {\dot{\bar\xi}}^i \\
\nonumber\!\!&-&\!\!\tilde{\mathcal{H}}^0_c-u\,\Omega-u^i\Omega_i-v\, 
\chi'-v^i\chi'_i-w \,\tau'-w^i\tau'_i\Big]\,, 
\end{eqnarray}
where
\begin{eqnarray}\label{B13}
\tilde{\mathcal{H}}^0_c&=&-\,\frac{d-2}{d-1}\,\omega^2+\frac{d-3}{4(d-2)}\,\omega_i\omega_i+\xi^k({\tilde
  h}_{,k}+\omega_k)+\frac{t}{d-1}\,\left({\tilde
  h}^l_{,\,l}-\omega-\tilde
\omega_{ll}\right)\\   
\nonumber 
&+&
\left(\tilde \omega_{ij}\,\tilde \omega_{ij}-\frac{1}{d-1}\,\tilde
\omega_{ii}\,\tilde \omega_{jj}-2\,\tilde\omega_{ij}{\tilde h}^i_{,j}+\frac{2}{d-1}\,\tilde
\omega_{kk}\,{\tilde h}^l_{,l}+{\tilde h}^i_{,j}{\tilde
  h}^j_{,i}-\frac{1}{d-1}\,{\tilde h}^k_{,k}{\tilde h}^l_{,l}\right)\\
\nonumber 
&-& 
\left(\frac{1}{2(d-2)}\,{\tilde
  h}^{ii}_{\,\,,j}\omega_j+\frac{1}{2}\,{\tilde
  h}^{jk}_{\,\,,i}{\tilde h}^{ik}_{\,\,,j}+\frac{1}{4(d-2)}\,{\tilde
  h}^{mm}_{\,\,\,\,,j}{\tilde h}^{nn}_{\,\,\,\,,j}   
-\frac{1}{4}\,{\tilde h}^{mn}_{\,\,\,\,,j}{\tilde h}^{mn}_{\,\,\,\,,j}\right)\,\,,
\end{eqnarray}
and
\begin{eqnarray}\label{5contrai}
\chi'&=&{\tilde h}^k_{,k}-\omega-\tilde\omega_{kk}\,,\\ \label{5contrai2}
\chi'_{i}&=&\tilde h_{,i}+\omega_i\,,\\ \label{5contrai3}
\tau'&=& {\tilde h}_{ij,ij}+\omega_{i,i}\,,\\ \label{5contrai4}
\tau'_i&=&2\,\tilde \omega_{kk,i}-2\,\tilde \omega_{ki,k}\,.
\end{eqnarray}
The action of eq. (\ref{action2}), with the Hamiltonian of eq. (\ref{B13}) and the
first class constraints of eqs. (\ref{5contrai})-(\ref{5contrai4}),
indeed coincide with the extended
action principle for the free spin two field theory on a flat
spacetime in first order form as developed in \cite{5Ghalati2007-2}. The tertiary
constraints $\tau'$ and $\tau'_i$ in 
fact contribute to the generator of the linearized diffeomorphism
transformation of the ``linerized'' affine connections
$\Gamma^\lambda_{\mu\nu}$ as found in \cite{5Ghalati2007-2}. 

\section{Summary and Conclusion}\label{IV} 
A major distinction between the Dirac Hamiltonian formulation of the
first order EH action as performed in \cite{5Ghalati2007-2} and the ADM
Hamiltonian formulation of the same action
\cite{5Arnowitt1959-1,5Arnowitt1959-2,5Arnowitt1960,5Arnowitt1960-2,5Arnowitt,5Faddeev1975,5Faddeev1982}        
is that in the latter all ``algebraic'' constraints are solved
in order to eliminate a number of fundamental fields from the action
at the Lagrangian level, while in the analysis of
\cite{5Ghalati2007-2} only those algebraic constraints which  
are second class (in the sense of the Dirac constraint formalism) are
used to eliminate fundamental fields; first class ``algebraic
constraints'' are treated according  to the Dirac constraint
formalism. This results in the appearance of 
tertiary first class constraints, and an unusual PB algebra of first
class constraints apparently different from the ADM algebra of the Hamiltonian and
momentum constraints $\mathcal{H}$ and $\mathcal{H}_i$. Therefore, it
is very important to compare 
the results of this novel Hamiltonian formulation with the usual ADM
formulation of the first order EH action. Such a comparison
remains obscure however, especially because of the different choices of
the canonical variables made in these formulations. 

The connection between this Hamiltonian formulation and the Faddeev and ADM
formulations was considered in this chapter, first using the method of
Faddeev and Jackiw \cite{5Antonio-Garcia,5Faddeev1988}, and then by proposing
tentative gauge constraints for the reduction of the formalism in the context of the Dirac
constraints method \cite{5Henneaux1992,5Sundermeyer}.
At first, the variables $(h, h^i, H^{ij})$ employed in \cite{5Ghalati2007-2} were canonically
transformed to $(h, h^i, q^{ij})$, $(h,h^i,\gamma_{ij})$,
$(N,N^i,\gamma_{ij})$ and $(\alpha,\alpha^i,\gamma_{ij})$. Upon the
first set of transformations, the tertiary constraint $\tau$ of eq. (\ref{ax55}) splits into
several terms as in eq. (\ref{5ax55}), some of which depend on the secondary
constraints. Therefore, the new choice of the tertiary constraint $\tilde 
\tau$ of eq. (\ref{5ax55-2}) is made possible and a great
simplification of the algebra of constraints 
occurs, as in eqs. (\ref{5ax24}-\ref{5ax67}). The secondary constraints commute with the tertiary
constraints as a result, and the tertiary constraints coincide with
the Hamiltonian and momentum constraints $C_0$ and $C_i$ of
eqs. (\ref{mhat1},\ref{mhat2}) of the Faddeev formulation
\cite{5Faddeev1975,5Faddeev1982}. The successive 
canonical transformations mentioned above were performed 
considering the new tertiary constraint $\tilde \tau$ rather
than $\tau$ as the tertiary constraint arising from the secondary constraint $\tilde \chi$. 
A choice of $(h, h^i, q^{ij})$, $(h,h^i,\gamma_{ij})$ or
$(\alpha,\alpha^i,\gamma_{ij})$ was demonstrated to be preferred to a choice of
$(h,h^i,H^{ij})$ or $(N,N^i,\gamma_{ij})$ as coordinates of the
Hamiltonian formulation, since the constraints take a especially simple form when
expressed in terms of the former sets of variables; the secondary first class
constraints depend only on the variables which are absent in the tertiary
constraints, and vice versa. This not only simplifies the task of 
determining the gauge transformations produced by the first class
constraints, but also reveals the unimportant
role of the subset of canonical variables $(h,h^i)$ or
$(\alpha,\alpha^i)$ in the formalism. More importantly, gauge fixing
of the extended Hamiltonian action becomes more transparent when the
former sets of variables are used. 

Considering the equations of motion arising from the 
Hamiltonian EH action when written in terms of $(h,h^i,q^{ij})$, we observe 
that there are no dynamical restrictions on the fields $h$ and $h^i$,
and thus they may be considered as Lagrange multiplier fields when
multiplied into the tertiary constraints $\tau$ and $\tau_i$. This was
illustrated in an alternative way by adding surface terms to the action
and transforming the secondary constraints $\tilde \chi$
and $\tilde \chi_i$ into the momenta conjugate to $h$ and $h^i$. The
necessary surface terms are equal to the surface terms added to the second order EH action by
Dirac \cite {5Dirac1958} in order to facilitate the task of a Hamiltonian formulation of
this action.  

When $(h,h^i,q^{ij})$ are used as coordinates, the gauge
transformation of the field $h$ generated by the first 
class constraints coincides with the diffeomorphism invariance
transformation of this field if the descriptor of the diffeomorphism
invariance has the particular dependence on the canonical variables and
the gauge functions of eqs. (\ref{trt},\ref{trt2}). Though our results
correspond only to the case where the gauge functions associated with the
tertiary constraints do not depend on 
the canonical variables, we expect that this feature is valid under more
general assumptions. The relationship between the gauge generator
and the descriptor of the diffeomorphism invariance has been considered in
\cite{5Bergmann,5Pons,5Salisbury}. Although we have only determined the
explicit form of the diffeomorphism invariance of $h$ in this chapter,
it is possible to find the gauge transformations of all other fields
from the formalism developed in the foregoing sections, thus 
the gauge transformations of the Christoffel symbols, as briefly
pointed out. In the ADM
approach, however, one needs to make use of the equations of motion
for the Christoffel symbols in order to determine their gauge
invariance. 

It is interesting to investigate if the Dirac quantization of the
above Hamiltonian formulations, in which first class constraints act as
operators, would produce results other than quantization of the
ADM action in which ``recduction'' is done before quantization. The
importance of this issue has been discussed in
\cite{5Antonio-Garcia,5Kuchar}.

\section{acknowledgments}
The author would like to thank K. Kargar, I. Khavkin and D.G.C. McKeon
and colleagues from the University of Western Ontario for the
enjoyment of numerous discussions. An unfinished collaboration with
Prof. McKeon on the gauge invariance of the action of eq. (\ref{action})
was helpful.

\pagebreak

\end{document}